\documentclass[oldstyle]{mtt}
\usepackage{caption}
\usepackage[font={rm,footnotesize},labelfont={rm,footnotesize}]{subcaption}

\usepackage[unicode]{hyperref}

\addtocounter{page}{61}

\udk{531.38+517.938.5}

\title[Атлас диаграмм обобщения 4-го класса особо замечательных движений]{АТЛАС ДИАГРАММ ОБОБЩЕНИЯ 4-ГО КЛАССА\\ОСОБО ЗАМЕЧАТЕЛЬНЫХ ДВИЖЕНИЙ АППЕЛЬРОТА\\НА ГИРОСТАТ В ДВОЙНОМ ПОЛЕ
}

\author{П.Е.~Рябов, Г.Е.~Смирнов, М.П.~Харламов}

\thanks{\hspace{-5mm}{Работа выполнена при финансовой поддержке РФФИ (грант № 10-01-00043).}}

\mtt{year=2012,number=42,received=28.08.12}

\address{Волгоградский филиал РАНХиГС, Россия}
\email{mharlamov@vags.ru}

\newcommand{\ds}{\displaystyle}
\newcommand{\diag}{\mathop{\rm diag}\nolimits}
\newcommand{\mam}{\mathcal{O}}
\newcommand{\bR}{\mathbb{R}}
\newcommand{\gs} {\geqslant}
\newcommand{\ls} {\leqslant}

\begin{document}

\maketitle

\begin{abstract}
В системе с двумя степенями свободы, которая является аналогом 4-го класса
Аппельрота для гиростата с условиями типа Ковалевской в двойном поле, решена задача классификации бифуркационных диаграмм. Построено разделяющее множество и дано доказательство его полноты. Представлены все преобразования, имеющие место в диаграммах. Результаты являются необходимым шагом в решении проблемы построения топологических инвариантов для интегрируемой системы Реймана--Семенова-Тян-Шанского с тремя степенями свободы.
\end{abstract}

\section*{Введение}
Исследуя вопрос о типах движений волчка С.В.\,Ко\-ва\-лев\-ской, Г.Г.\,Аппельрот выделил четыре класса движений, которые он назвал {\it особо замечательными}. В них остается постоянной одна из разделенных переменных Ковалевской. Как выяснилось позже, классы Аппельрота полностью исчерпывают критическое множество отображения момента и, следовательно, порождают все бифуркации интегральных многообразий. Случай Ковалевской распространен на гиростат в работах Х.М.Яхья\, и на гиростат в двойном поле в работе \cite{ReySem}. При этом первые три класса Аппельрота претерпели принципиальные изменения, а последний получил естественное обобщение \cite{KhND07}. Отметим, что 4-й класс Аппельрота является частным случаем классического решения Бобылева\,--\,Стеклова, а его аналог для гиростата в поле силы тяжести включен в более общее семейство решений, найденное в~\cite{PVLect}.

Критическое множество отображения момента гиростата с условиями типа Ковалевской в двойном поле состоит из многообразий четной размерности ниже 6, на которых индуцируются почти всюду гамильтоновы системы с меньшим числом степеней свободы. Одной из таких {\it критических подсистем} и является обобщение 4-го класса Аппельрота, данное в работе \cite{KhND07}. При отсутствии гиростатического момента (для {\it волчка} в двойном поле) соответствующее многообразие указано в \cite{Kh34}. Индуцированная на нем система полностью исследована в работах \cite{KhShRCD06, Kh38, Kh40}.

Опишем систему, изучаемую в настоящей работе. Рассмотрим гиростат в двойном поле и обозначим через $\mathbf{M}$ и ${\bs \omega}$ векторы кинетического момента и угловой скорости, так что $\mathbf{M}= \mathbf{I}{\bs \omega} +{\bs \lambda}$. Далее для аналога случая Ковалевской полагаем $\mathbf{I}=\diag\{2,2,1\}$, ${\bs \lambda}=(0,0,\lambda)$. Пусть ${\boldsymbol\alpha},{\boldsymbol\beta}$ -- характеристические векторы силовых полей, а центры приложения полей лежат в экваториальной плоскости эллипсоида инерции. Без ограничения общности можно считать, что радиус-векторы центров приложения составляют ортонормированную пару (и, следовательно, могут быть взяты в качестве ортов первых двух главных осей инерции), а векторы ${\boldsymbol\alpha},{\boldsymbol\beta}$ взаимно ортогональны~\cite{Kh34}. В силу последнего, фазовое пространство системы с тремя степенями свободы задается в $\bR^9({\boldsymbol\omega},{\boldsymbol\alpha},{\boldsymbol\beta})$ геометрическими интегралами
$$
{\boldsymbol\alpha}^2=a^2, \quad {\boldsymbol\beta}^2=b^2, \quad {\boldsymbol\alpha} \cdot {\boldsymbol\beta} =0 \quad (a \gs b \gs 0, \quad  a^2+b^2 \ne 0).
$$
В случае ${a\gs b>0}$ это пространство шестимерно, а в случае ${a > b>0}$ в задаче отсутствует циклический интеграл.
Рассмотрим функцию
$$
\Phi=\frac{{\bs M}\cdot {\boldsymbol\alpha}}{\omega_1}-\frac{{\bs M}\cdot {\boldsymbol\beta}}{\omega_2}
$$
и систему инвариантных соотношений $\{\Phi =0, \,{\dot \Phi} =0\}$.
Замыкание множества точек, заданного этой системой, инвариантно относительно уравнений движения и является почти всюду четырехмерным многообразием, симплектическим относительно ограничения исходной симплектической структуры. По аналогии с работой \cite{KhND07} обозначим это замыкание вместе с индуцированной на нем (почти) гамильтоновой системой с двумя степенями свободы через $\mam$ и будем называть {\it основной} системой (для настоящей работы). В \cite{KhND07} указана связь этой системы с 4-м классом Аппельрота. К настоящему моменту для основной системы имеются следующие результаты. В работе \cite{Kh37} найдены все так называемые ``особые периодические движения'' (далее, сокращенно, -- ОПД), т.е. траектории, состоящие из критических точек интегрального отображения ранга 1 (среди них есть и непериодические, но асимптотические к положениям равновесия гиростата). На этих движениях исходная система интегрируется в эллиптических функциях. Условия вещественности соответствующих решений (т.е. условия существования ОПД в пространстве параметров) найдены в \cite{KhIISmir}. В работе \cite{c:main} в терминах констант пары функционально независимых интегралов вычислен показатель {\it внешнего} типа для всех критических точек интегрального отображения, что, в частности, дает локальную классификацию регулярных точек системы $\mam$, рассматриваемых как особенность ранга~2 в системе с тремя степенями свободы.
Настоящая публикация использует результаты и формулы работ \cite{KhND07,Kh37,c:main}. Система $\mam$ соответствует {\it первой критической подсистеме $\mathcal{M}_1$} работы \cite{c:main}.

Согласно \cite{KhND07} в качестве пары коммутирующих интегралов на $\mam$ можно взять функции $H,S$:
\begin{equation}\label{eq:1}
\begin{array}{l}
\displaystyle{H = \omega_1^2+\omega_2^2+\frac{1}{2}\omega_3^2- \alpha_1 -\beta_2},\\[3mm]
\displaystyle S=-\frac{2 \alpha_1 \omega_1^2+2 (\alpha_2+\beta_1) \omega_1 \omega_2+2 \beta_2 \omega_2^2 +(\alpha_3\omega_1+\beta_3\omega_2)(\omega_3+\lambda) }{2(\omega_1^2+\omega_2^2)}.
\end{array}
\end{equation}
Последний интеграл выбран не случайно. Его константа $s$ в предельной классической задаче (${\bs \beta}=0$, $\lambda=0$) есть значение переменной разделения С.В.\,Ковалевской, остающейся постоянной в соответствующих движениях. В работе \cite{KhND07} указана также связь этого значения со спектральным параметром представления Лакса.

Константы $k,g$ двух общих интегралов исходной системы $K$ и $G$ в точках подсистемы $\mam$ выражаются в терминах констант $h,s$ интегралов \eqref{eq:1}, формируя первую бифуркационную поверхность $\Pi_1$ в $\bR^3(h,k,g)$:
\begin{equation} \label{eq:2}
\Pi_1: \left\{ \begin{array}{l} \displaystyle{k =
p^2+(h-\frac{\lambda^2}{2})^2-4(h-\frac{\lambda^2}{2})s+3s^2-
\frac{p^4-r^4}{4s^2},} \\[3mm]
\displaystyle{g=(h-\frac{\lambda^2}{2}-s)s^2+\frac{p^4-r^4}{4s},
\qquad s \in {\mathbb{R}}\backslash\{0\}.}
\end{array} \right.
\end{equation}
Здесь и далее наряду с параметрами $a,b$ используются вспомогательные величины $p \gs r \gs 0$, такие что $p^2=a^2+b^2$, $r^2=a^2-b^2$.

Нам понадобятся и уравнения второй бифуркационной поверхности $\Pi_2$~-- образа другой критической подсистемы $\mathcal{M}_2$ под действием отображения момента \cite{c:main}:
\begin{equation} \label{eq:3}
\Pi_2: \left\{ \begin{array}{l} \displaystyle{k = -
2\lambda^2(h-\frac{\lambda^2}{2}-2s)-\lambda^4+ \frac{r^4}
{4s^2},} \\[3mm]
\displaystyle{g=\frac{1}{2}p^2(h+\frac{\lambda^2}{2})-\lambda^2 s^2-\frac{r^4}{4s},
\qquad s \in {\mathbb{R}}\backslash\{0\}.}
\end{array} \right.
\end{equation}
Здесь $s$ -- постоянная соответствующего частного интеграла для $\mathcal{M}_2$.

При классификации критических точек в системе с тремя степенями свободы по возникающим в их окрестности слоениям Лиувилля в предположении отсутствия так называемых расщепляющихся атомов разделяющим служит множество критических точек ранга~0 и 1, а также вырожденных критических точек ранга~2 (в системе с $n$ степенями свободы, соответственно, -- критических точек ранга до $n-2$ включительно и вырожденных точек ранга $n-1$). Это множество точек называется {\it ключевым} множеством системы \cite{KhRyabUdgu2011}.

{\bf Определение} \cite{KhRyabUdgu2011}. Пусть у критической подсистемы заданы два независимых почти всюду интеграла $\phi$ и $\psi$.
Назовем $(\phi,\psi)$-диаграммой критической подсистемы образ принадлежащих ей точек ключевого множества при отображении $\phi{\times}\psi$.

Поставим следующие задачи для основной системы $\mam$:

-- получить аналитическое описание $(S,H)$-диаграммы системы и указать способ ее построения;

-- найти атлас (разделяющее множество) в пространстве параметров $a,b,\lambda$, классифицирующий такие диаграммы;

-- указать перестройки, происходящие в диаграммах при пересечении раз\-деляющих кривых в составе атласа.

\section{Ключевое множество}\label{sec1}
В неприводимом случае $b>0$ критических точек ранга 0 (положений равновесия) в рассматриваемой задаче четыре:
\begin{equation}\label{eq:4}
c_k: \; {\boldsymbol\omega}={\boldsymbol 0}, \quad  {\boldsymbol
\alpha} = (\varepsilon_1  a ,\, 0,\, 0), \quad {\boldsymbol \beta}=
(0,\, \varepsilon_2 b, \, 0).
\end{equation}
Здесь $\varepsilon_1^2=\varepsilon_2^2=1$ и $k=1,\ldots,4.$ Упорядочим $c_k$ по возрастанию $h$:
\begin{equation}\label{eq:5}
    h_1=-a-b, \quad h_2=-a+b, \quad h_3=a-b, \quad h_4=a+b.
\end{equation}

Критические точки ранга 1 организованы в две различных группы движений. Первая группа~-- маятниковые движения, в которых осью маятников служит ось динамической симметрии $O{\bf e}_3$, расположенная ортогонально обоим силовым полям. В зависимости от ориентации образовавшегося триэдра $({\boldsymbol \alpha},{\boldsymbol \beta}, {\bf e}_3)$ соответствующие многообразия, сотканные из маятниковых траекторий,
обозначаются через $\mathcal{M}_{\pm}$ \cite{c:main}. Очевидно, $c_1,c_4 \in \mathcal{M}_+$, $c_2,c_3 \in \mathcal{M}_-$ и точки $c_k$ служат бифуркационными внутри семейств маятников. Точки $c_k$ принадлежат основной системе $\mam$ за одним исключением, а именно, точка $c_4$ самого верхнего положения равновесия не принадлежит $\mam$, если значение энергии $h=a+b$ лежит в интервале
\begin{equation}\label{eq:6}
\frac{\lambda^2}{2}-2\sqrt{a b} < h < \frac{\lambda^2}{2}+2\sqrt{a b}.
\end{equation}
Множество $\mathcal{M}_-$ целиком содержится в $\mam$, а $\mathcal{M}_+$ пересекается с подсистемой $\mam$ только по движениям со значениями энергии за пределами интервала~\eqref{eq:6}. Множества $\mathcal{M}_\pm$ при отображении момента переходят в прямые $\Pi_\pm$. Соответствующие формулы приведены в \cite{c:main}.

Вторая группа критических точек ранга 1 составлена из ``особых периодических движений'', заданных уравнениями (8), (27), (29), (30) работы \cite{Kh37}. Для экономии места эти уравнения здесь не повторяем и будем ссылаться на них как на уравнения (ОПД). Система уравнений (ОПД) выражает все фазовые переменные через одну переменную $w$, физические параметры $a,b,\lambda$ и два вспомогательных параметра $\sigma, u$, связанных соотношением
\begin{equation}\label{eq:7}
\begin{array}{l}
\lambda^2(\lambda^2+\sigma)^2 u^5+(\lambda^2+\sigma) [2p^2\lambda^4-(\lambda^2+\sigma)^3 \sigma]\sigma u^4 + \\
\qquad +r^4\lambda^6\sigma^2 u^3+ 2 r^4 \lambda^4 \sigma^4(\lambda^2+\sigma)^2u^2-r^8\lambda^8 \sigma^6=0.
\end{array}
\end{equation}
Изменение переменной $w(t)$ определяется дифференциальным уравнением
\begin{equation} \label{eq:8}
\displaystyle{\big(\frac{dw}{dt}
\big)^2=-\frac{\lambda^2}{4\sigma^2} P_+ (w) P_- (w),}
\end{equation}
где
\begin{equation}\notag
\displaystyle{P_{\pm}(w) = w^2+2\sigma^2\frac{u \pm
r^2\lambda^2}{\lambda^2 u}w +
\frac{\sigma[u^3-(\lambda^2+\sigma)\sigma^2u^2+r^4\lambda^4
\sigma^3]}{(\lambda^2+\sigma)\lambda^2u^2}.}
\end{equation}
Условия существования решений уравнений (ОПД) найдены в работе \cite{KhIISmir}.

Наконец, условия вырожденности получены в работе \cite{c:main}. В частности, для регулярных точек системы $\mam$ имеет место теорема.
\begin{theorem}[\cite{c:main}]
{\it Регулярные точки системы $\mam$, рассматриваемые как критические точки ранга~$2$ в объемлющей системе с тремя степенями свободы, имеют тип, определенный знаком выражения
$\mu=s \, F_C(s,h) F_T(s,h)$, где $h,s$ -- значения интегралов \eqref{eq:1},
$$
F_C(s,h)=3 s^4-2(h-\frac{\lambda^2}{2})s^3+a^2 b^2, \quad F_T(s,h)=2s^2-2(h+\frac{\lambda^2}{2})s+a^2+b^2.
$$
Критические точки имеют тип ``центр'' при ${\mu<0}$ и тип ``седло'' при ${\mu>0}$. При ${\mu=0}$ критические точки вырождены. В пространстве констант первых интегралов множество вырожденных критических точек отвечает ребру возврата поверхности $\Pi_1$
\begin{equation}\label{eq:9}
    F_C(s,h)=0
\end{equation}
и линии
\begin{equation}\label{eq:10}
    F_T(s,h)=0
\end{equation}
касания поверхности $\Pi_1$ с другой бифуркационной поверхностью $\Pi_2$, заданной уравнениями {\rm \eqref{eq:3}}.}
\end{theorem}

\section{Подмножества в диаграмме}
По определению диаграммы и в силу результатов раздела \ref{sec1} диаграмма состоит из образов на плоскости $(s,h)$ следующих множеств:

1) положений равновесия;

2) маятниковых движений $\mathcal{M}_+$;

3) маятниковых движений $\mathcal{M}_-$;

4) особых периодических движений, определенных уравнениями (ОПД);

5) критических торов из точек ранга 2, отвечающих линии касания поверхностей $\Pi_1, \Pi_2$;

6) критических торов из точек ранга 2, отвечающих ребру возврата поверхности $\Pi_1$.

Обозначим образы этих множеств на плоскости $(s,h)$ через $D_0$, $D_+$, $D_-$, $D_1$, $D_T$, $D_C$ соответственно. Уравнения всех этих составляющих диаграммы, кроме $D_1$, записываются явно.

Обозначим
\begin{equation}\notag
\begin{array}{c}
   \ds{Q_{\pm}(x)=\sqrt{\left(x-\frac{\lambda^2}{2}\right)^2 \mp 4 a b}}\ \gs 0, \\[3mm]
   \ds{\phi^+_{1,2}(x)=\frac{1}{2}\left[x-\frac{\lambda^2}{2} \mp Q_+(x) \right]},\qquad \ds{\phi^-_{1,2}(x)=\frac{1}{2}\left[x-\frac{\lambda^2}{2} \mp Q_-(x) \right]},
\end{array}
\end{equation}
и пусть для $j=1,2$
\begin{equation}\notag
\begin{array}{c}
    s_{1j}=\phi^+_j(-a-b), \quad s_{2j}=\phi^-_j(-a+b), \quad s_{3j}=\phi^-_j(a-b), \quad s_{4j}=\phi^+_j(a+b).
\end{array}
\end{equation}
Множество $D_0$ состоит из восьми или шести (значения $s_{4j}$ не всегда вещественны) точек, отвечающих положениям равновесия~\eqref{eq:4} со значениями энергии \eqref{eq:5}\footnote{Интеграл $S$ неоднозначно определен на маятниковых движениях и, в частности, в положениях равновесия.}:
\begin{equation}\label{eq:11}
P_{ij}(s_{ij}, h_i) \qquad  (i=1,\ldots,4, \;j=1,2).
\end{equation}
Множество $D_+$ состоит из ветвей гиперболы (для отрицательных $s$ -- сегмент ветви, ограниченный точками $P_{11},P_{12}$):
\begin{equation}\notag
  \ds{h=\frac{\lambda^2}{2}+s+\frac{ab}{s}}, \qquad s \in [s_{11},s_{12}] \cup  (0,+\infty).
\end{equation}
Для обратных зависимостей имеем
\begin{equation}\notag
    \ds{s=\phi^+_{1,2}(h)}, \qquad h \in [-a-b, -2\sqrt{a b}+\frac{\lambda^2}{2}]\cup [2\sqrt{a b}+\frac{\lambda^2}{2},+\infty).
\end{equation}
Множество $D_-$ состоит из сегментов ветвей гиперболы, ограниченных снизу по $h$ точками $P_{21},P_{22}$:
\begin{equation}\notag
  \ds{h=\frac{\lambda^2}{2}+s - \frac{a b}{s}}, \qquad s \in [s_{21},0)  \cup  [s_{22},+\infty).
\end{equation}
Очевидно, что $s_{31}\in (s_{21},0), s_{32}\in (s_{22},+\infty)$, поэтому точки $P_{3j}$ также учтены.
Для обратных зависимостей
\begin{equation}\notag
  \ds{s=\phi^-_{1,2}(h)}, \qquad h \in [-a+b, +\infty).
\end{equation}
Множество $D_T$ описывается уравнением \eqref{eq:10}, а множество $D_C$ -- уравнением~\eqref{eq:9}. Вопрос об области изменения $s$ в этих случаях пока открыт.

Неявное уравнение множества $D_1$ получим так. Из уравнений {\rm \eqref{eq:2}}, {\rm \eqref{eq:3}} запишем условия пересечения поверхностей $\Pi_1,\Pi_2$~-- два уравнения для совпадения координат $g,k$ при одном и том же $h$, но произвольных различных значениях параметра $s$, оставив обозначение $s$ для $\Pi_1$, а для $\Pi_2$ обозначив этот параметр, например, через $t$. Вычислим результант левых частей этих уравнений по $t$. Получим уравнение вида $F_T^2(s,h) \Phi(s,h)=0$,
где $\Phi(s,h)$ -- многочлен степени 4 по $h$ и степени 12 по $s$ с коэффициентами, полиномиальными по $a,b,\lambda$. Его несложно выписать с помощью компьютерной алгебры, поэтому для краткости соответствующее выражение здесь не приводим. Очевидно, сомножитель $F_T^2(s,h)$ отвечает за касание поверхностей, а уравнением пересечения в точке общего положения является
$\Phi(s,h)=0$.
Однако для визуализации множества $D_1$ это уравнение непригодно, так как, определив удовлетворяющую ему точку $(s,h)$, мы не имеем способа установить, является ли она допустимой, т.е. отвечает ли она некоторому вещественному решению уравнений (ОПД), например, условиям, полученным в \cite{KhIISmir}. В связи с этим предлагается следующий численный алгоритм построения $D_1$.

Фиксируем физические параметры $a,b,\lambda$. Сразу же отметим, что всегда можно считать, что
\begin{equation}\label{eq:12}
    a=1, \qquad 0 \ls b \ls 1, \qquad \lambda \gs 0,
\end{equation}
так как $a$ можно взять за единицу измерения длин векторов ${\boldsymbol \alpha}, {\boldsymbol \beta}$ и выбрать подходящее направление подвижных осей.

Выбираем промежуток значений энергии $[h_{\rm min},h_{\rm max}]$, на котором лежат все особые точки диаграммы. Ниже эти точки названы узловыми. Их конечное число и все они эффективно вычисляются. Очевидно, что в качестве $h_{\rm min}$ всегда нужно брать абсолютный минимум энергии $h_{\rm min}=h_1=-a-b$. Значение $h_{\rm max}$ вычисляется как наибольшая из $h$-координат узловых точек, увеличенная так, чтобы увидеть поведение кривых в окрестности наивысшей узловой точки.
Выражение для $h$ в произвольной точке ОПД получим из формулы (35) работы \cite{Kh37} в виде $h=\varphi(u,\sigma)$. Из этого уравнения и уравнения \eqref{eq:7} получим одно уравнение $\Psi(h,\sigma)=0$, связывающее $h,\sigma$, и выражение для общего корня $u$ в виде $u=\theta(h,\sigma)$, которое, хотя и является весьма громоздким, но легко выписывается средствами компьютерной алгебры в виде дробно-рациональной функции всех входящих параметров. Меняя $h$ с определенным шагом на отрезке $[h_{\rm min},h_{\rm max}]$, находим для каждого $h$ все решения $\sigma$ уравнения $\Psi(h,\sigma)=0$. Для каждой найденной пары $(h,\sigma)$ вычисляем $u=\theta(h,\sigma)$ и проверяем наличие положительного корня по $w$ многочлена $P_+(w; u,\sigma)P_-(w; u,\sigma)$, что является необходимым и достаточным условием существования вещественного решения уравнений (ОПД), так как по своему определению переменная $w$ в \eqref{eq:8} неотрицательна. В случае наличия положительного корня координата $s$ соответствующей точки из $D_1$ находится из формулы (40) работы \cite{Kh37}.
В результате такой процедуры множество $D_1$ получается в виде большого массива точек, что даже на достаточно мощных компьютерах не позволяет наблюдать перестройки в диаграмме, меняя динамически физические параметры. Эту проблему удается обойти с помощью алгоритмов замены точек кривыми, использующих полученные в работе \cite{c:main} формулы для вычисления типов точек множества $D_1$ как критических точек ранга~1. Описание таких алгоритмов выходит за рамки настоящей статьи.

\section{Узловые точки диаграммы}\label{sec3}
В этом разделе указаны особые точ\-ки диаграммы, бифуркации в множестве которых порождают перестройки самой диаграммы.
Все предложения проверяются прямым вычислением и потому приводятся без доказательства.
Отметим, что в рамках настоящей работы не удалось найти способ установить допустимые сегменты $D_T,D_C$ на кривой касания поверхностей и на ребре возврата поверхности $\Pi_1$, то есть множества тех точек кривых, заданных уравнениями \eqref{eq:10}, \eqref{eq:9}, в прообразе которых существуют траектории из $\mam$.
Здесь и далее мы будем исходить из гипотезы отсутствия минимальных торов.

{\bf Гипотеза.} {\it При каждом фиксированном $s$ на множестве ${\mam \cap \{S=s\}}$ наименьшее значение $h$ достигается в критической точке ранга не выше~$1$.}

В частности, это означает, что точки общего положения на $D_T$ и $D_C$ не могут выступать нижней границей допустимой области по $h$.
Следующее утверждение описывает пересечения $D_C$ и $D_T$ с кривыми $D_\pm$ и позволяет определить границу $D_T$.
\begin{proposition}\label{propn1}
{\it $1)$ Кривая $D_T$ имеет с каждой из кривых $D_-$ и $D_+$ ровно по одной общей точке}
\clearpage
\begin{equation}\label{eq:13}
\begin{array}{l}
P_-=D_T\cap D_-=\left\{\ds{s=\frac{(a + b)^2}{2 \lambda^2}}, \quad \displaystyle{h}=\displaystyle{\frac{(a+b)^4+(a- b)^2 \lambda^4}{2(a+b)^2 \lambda^2}}\right\},\\
P_+=D_T\cap D_+=\left\{\ds{s= \frac{(a - b)^2}{2 \lambda^2}}, \quad \displaystyle{h}=\displaystyle{\frac{(a-b)^4+(a+b)^2 \lambda^4}{2(a-b)^2 \lambda^2}}\right\}.
\end{array}
\end{equation}

{\it
$2)$ Множество $D_T$ на кривой \eqref{eq:10} задано условием
$\displaystyle{\ds{s\in (0, \frac{(a + b)^2}{2 \lambda^2}],}}$
где правая граница определяется точкой $P_-$ пересечения с $D_-$.
}

{\it
$3)$ Кривая $D_C$ не имеет общих точек с $D_-$ и ровно две общих точки с $D_+$, а именно, изолированную точку
\begin{equation}\label{eq:14}
s=-\sqrt{ab},\qquad \displaystyle{h=\frac{\lambda^2}{2}-2\sqrt{ab}}
\end{equation}
и точку касания
\begin{equation}\label{eq:15}
s=\sqrt{ab},\qquad \displaystyle{h=\frac{\lambda^2}{2}+2\sqrt{ab}}.
\end{equation}
Последняя точка не является граничной для $D_C$.
}
\end{proposition}

Отметим, что если рассматривать кривую \eqref{eq:9} без учета условий существования вещественных решений, то точка \eqref{eq:14} также оказывается точкой касания этой кривой с  $D_+$. Однако легко показать, что в целом при отрицательных $s$ кривая \eqref{eq:9} лежит ниже $D_+$. Поэтому из принятой гипотезы следует, что ее точки для $s<0$, отличные от  \eqref{eq:14}, недопустимы.

Рассмотрим пересечение $D_C$ с $D_T$.
\begin{proposition}\label{propn2}
{\it
При условии
\begin{equation}\label{eq:16}
    \lambda^4 < \ds{\frac{a^6-33 a^4 b^2-33 a^2 b^4+b^6+(a^4+14a^2b^2+b^4)^{3/2}}{54a^2b^2}}
\end{equation}
кривые $D_T,D_C$ имеют ровно две общие точки, определяемые системой
\begin{equation}\label{eq:17}
    (a^2-s^2)(s^2-b^2)-2\lambda ^2 s^3=0, \qquad s >0, \qquad \ds{h=\frac{2 s^2 - s \lambda^2+a^2+b^2}{2 s}}.
\end{equation}
Если в \eqref{eq:16} неравенство заменить равенством, то эти точки сливаются, образуя точку касания кривых $D_T$ и $D_C$. При обратном строгом неравенстве в \eqref{eq:16} кривые $D_T,D_C$ общих точек не имеют.
}
\end{proposition}

Очевидно, и эти точки пересечения не могут служить граничными для $D_C$. Следовательно, граница $D_C$ должна лежать в $D_1$. Перейдем к рассмотрению пересечений $D_1$ с другими подмножествами диаграммы.

\begin{proposition}\label{propn3}
{\it
$1)$ Кривая $D_1$ имеет с каждой из кривых $D_-$ и $D_+$ в качестве общих точек образы положений равновесия -- точки \eqref{eq:11}, а также ровно по одной общей точке касания
\clearpage
\begin{equation}\label{eq:18}
\begin{array}{l}
P_-^*=\left\{ \ds{s= -\frac{2 a b \lambda^2}{(a + b)^2}}, \quad \displaystyle{h}=\displaystyle{\frac{(a+b)^4+(a- b)^2 \lambda^4}{2(a+b)^2 \lambda^2}}\right\}\in D_1\cap D_-,\\
P_+^*=\left\{ \ds{s= \phantom{-}\frac{2 a b \lambda^2}{(a - b)^2}}, \quad \displaystyle{h}=\displaystyle{\frac{(a-b)^4+(a+b)^2 \lambda^4}{2(a-b)^2 \lambda^2}}\right\}\in D_1\cap D_+.
\end{array}
\end{equation}

$2)$ Общими точками кривых $D_1, D_T$ служат общие точки кривых $D_T,D_C$ {\rm (}которые являются, таким образом, тройными точками диаграммы{\rm )} и, при условии $\lambda \ls (a+b)/\sqrt{a-b}$, единственная точка касания $D_1$ и $D_T$, заданная координатами
\begin{equation}\label{eq:19}
P_T:\quad s=\frac{r^{4/3}}{2 \lambda^{2/3}}, \qquad h=\frac{ r^{8/3} + 2 p^2 \lambda^{4/3} - r^{4/3} \lambda^{8/3}}{2 r^{4/3} \lambda^{2/3}}.
\end{equation}
Других общих точек у множеств $D_T, D_1$ нет.
}
\end{proposition}

Отметим, что точка \eqref{eq:19} соответствует случаю устранимой особенности $\sigma = -\lambda^2$ в уравнениях (ОПД). Подчеркнем также, что точки \eqref{eq:13}, \eqref{eq:14}, \eqref{eq:15}, \eqref{eq:18} отвечают вырожденным маятниковым движениям \cite{c:main}.

\begin{proposition}\label{propn4}
{\it
Точки пересечения $D_C$ и $D_1$, отличные от тройных точек, находятся из системы
\begin{equation}\label{eq:20}
    F_C(s,h)=0, \qquad \Phi_C(s)=0, \qquad s>0, \qquad h > -a-b,
\end{equation}
где
\begin{equation}\notag
\begin{array}{l}
\Phi_C(s)=  [a^4 b^4 - 6 a^2 b^2 s^4 + 4 (a^2 + b^2) s^6 - 3 s^8]^2 + \\
\qquad + 2 [3 a^6 b^6 + 3 a^4 b^4 s^4 - 20 a^2 b^2 (a^2 + b^2) s^6 + 57 a^2 b^2 s^8 - 12 (a^2 + b^2) s^{10} +\\
\qquad + s^{12}] s^3 \lambda^2 + 4  [3 a^4 b^4 + 15 a^2 b^2 s^4 - (a^2 + b^2) s^6] s^6 \lambda^4 + 8 a^2 b^2 s^9 \lambda^6.
\end{array}
\end{equation}
}
\end{proposition}

Теперь в силу принятой выше гипотезы допустимый промежуток изменения параметра $s$ на кривой $D_C$ определяется следующим образом.
\begin{proposition}\label{propn5}
{\it
Кривая $D_C$ определяется уравнением \eqref{eq:9} со значениями переменной $s$
\begin{equation}\notag
s\in \left\{-\sqrt{a b}\,\right\} \cup (0, s_*].
\end{equation}
Здесь $s_*=s_*(\lambda,a,b)$ -- наибольший вещественный корень многочлена $\Phi_C(s)$. Такой корень существует и положителен при всех $\lambda$.
}
\end{proposition}

Поскольку кривые $D_{\pm},D_T,D_C$ имеют весьма простую структуру без особых точек, а все кратные точки уже изучены, для описания особых точек диаграммы остается найти точки возврата кривых в составе $D_1$. Они соответствуют пересечению поверхности $\Pi_1$ с ребром возврата поверхности $\Pi_2$.

\begin{proposition}\label{propn6}
{\it
Для того чтобы точка $(s,h)$, найденная из уравнений множества $D_1$, была допустимой точкой возврата $D_1$, необходимо и достаточно выполнение следующих условий{\rm :}

$1)$ $h \gs -a-b;$

$2)$ пара $(s,h)$ удовлетворяет системе уравнений
\begin{equation}\label{eq:21}
    \begin{array}{l}
       (2s^2-p^2)^2-r^4+3r^{4/3}\lambda^{4/3}s^2+2\lambda^2 s^3=0, \\
       \ds{h=\frac{4 s^4+2 \lambda^2 s^3-(3 \lambda^{2/3} r^{8/3}-\lambda^2 p^2 ) s-(p^4-r^4)}{2s(2s^2-p^2)}};
     \end{array}
\end{equation}

$3)$ для заданного $s$ среди значений $\sigma =-\lambda^2\pm \sqrt{r^{4/3}\lambda^{4/3}-2s \lambda^2}$, $u=-r^{4/3}\lambda^{4/3}\sigma$
найдутся такие, что многочлен $P_+(w; u,\sigma)P_-(w; u,\sigma)$ в правой части уравнения {\rm \eqref{eq:8}} имеет хотя бы один положительный корень.
}
\end{proposition}

Следующие отмеченные выше точки будем называть {\it узловыми}:

-- образы \eqref{eq:11} положений равновесия, они же -- точки трансверсального пересечения $D_1$ c $D_\pm${\rm ;}

-- образы вырожденных маятников, они же -- точки \eqref{eq:13} пересечения $D_T$ c $D_\pm${\rm ;}

-- точки \eqref{eq:14}, \eqref{eq:15}  касания $D_C$ c $D_+$, они же -- точки экстремума $h$ на $D_+${\rm ;}

-- две тройных точки \eqref{eq:17} пересечения $D_T$, $D_C$, $D_1${\rm ;}

-- образы вырожденных маятников для дуального значения $s$, они же -- точки \eqref{eq:18} касания $D_1$ c $D_\pm${\rm ;}

-- точка \eqref{eq:19} касания $D_1$ с $D_T${\rm ;}

-- точки \eqref{eq:20} пересечения $D_1$ c $D_C${\rm ;}

-- точки \eqref{eq:21} возврата кривых в составе $D_1$.

\def\T{T}
\def\P{R}
\section{Бифуркации семейства диаграмм}
Рассмотрим семейство определенных выше $(S,H)$-диаграмм основной системы, зависящих от набора параметров
\begin{equation}\notag
    c=(\lambda,a,b), \qquad \lambda\gs 0, \qquad 0<b<a.
\end{equation}
Обозначим произвольную диаграмму семейства через $\T=\T(c)$. При каждом $c$ в диаграмме выделен конечный набор точек, названных узловыми. Обозначим это конечное множество через $\P=\P(c)$.

Под бифуркациями плоских диаграмм по векторному параметру $c\in \bR^k$ обычно понимают нарушение локальной тривиальности расслоения-проекции ${\mathop{\rm pr}\nolimits: U \to \bR^k}$, где $U$ -- сколь угодно малая окрестность множества $\cup_c (\T(c),c)$ в $\bR^2{\times}\bR^k$. Здесь нужны аккуратные определения, поскольку диаграмма не является гладким многообразием. Однако в нашем случае диаграмма такова, что качественные перестройки ее структуры полностью определены перестройками в конечном множестве узловых точек.
А бифуркация в семействе множеств $\P(c)$, снабженных дискретной топологией, есть изменение количества точек в $\P(c)$, которое происходит либо на границе условий существования той или иной точки, либо при возникновении в множестве $\P(c)$ кратной точки. В силу различного вида узловых точек и принадлежности их различным кривым в составе диаграммы, необходимо описать явный и достаточно простой алгоритм перебора вариантов, гарантирующий нахождение всех необходимых и достаточных условий бифуркации.

Итак, пусть задано множество $\T(c)$ и в нем конечный набор точек $\P(c)$, называемых узловыми. Пусть имеется представление вида
$\T=\cup \T_i$, $\P=\cup \P_i$, где $\P_i \subset \T_i$.
Множества $\T_i$ назовем для краткости {\it слоями}. Предположим, что все точки пересечения двух слоев $\T_{ij}=\T_i \cap \T_j$ являются узловыми точками для одного из этих слоев. Поскольку порядок в совокупности множеств $\T_i$ по номеру $i$ вводится условно, то и отнесение точки $p\in \T_{ij}$ к тому или иному из подмножеств $\P_i, \P_j$ условно. Его удобно осуществлять по мере рассмотрения множеств $\T_i$ при первом появлении пересечения. Обозначим через $\P^0_i$ ту часть множества узловых точек $\P_i$, которая не попала в пересечения $\T_{ij}$ при $j<i$. При этом может оказаться, что при некоторых $c$ какие-то точки $\P^0_i(c)$ все же попадают в предыдущие множества. Такие значения параметра по определению считаются бифуркационными (разделяющими).

Дальнейший алгоритм состоит в следующем. Для первого множества $\T_1$ по определению полагаем $\P_1=\P^0_1$ и исследуем возможные бифуркации (появления кратных точек) внутри $\P^0_1$. Для множества $\T_i$ с $i>1$ исследуем: 1)~бифуркации внутри множеств $\cup_{j<i}\T_{ij}$; 2)~бифуркации внутри $\P^0_i$; 3)~случаи наличия непустых пересечений $(\cup_{j<i}\T_{ij}) \cap \P^0_i$.

В нашей задаче все узловые точки описаны в разделе~\ref{sec3}. Множества, формирующие диаграмму, можно занумеровать, например, в таком порядке:
$$
\T_1= D_+ \cup D_-, \quad \T_2=D_T, \quad \T_3=D_C, \quad \T_4=D_1.
$$
Тогда
$$
\begin{array}{l}
  \P_1=\P^0_1 = D_0, \qquad \P_2 = D_T \cap (D_+ \cup D_-), \\
  \P_3=(D_C \cap D_T) \cup (D_C \cap (D_+ \cup D_-)) \backslash (\P_1 \cup \P_2), \\
  \P_4 = \P^0_4 \cup (D_1  \cap (D_+ \cup D_-) ) \cup (D_1  \cap D_T) \cup (D_1 \cap D_C) \backslash (\P_1 \cup \P_2 \cup \P_3),
\end{array}
$$
где $\P^0_4$ -- точки возврата на кривых $D_1$. Бифуркации в $\cup_i \P_i$ находятся с использованием утверждений раздела~\ref{sec3}, которые описывают систему узловых точек именно в такой последовательности. В результате получаем следующее описание разделяющего множества.
\begin{theorem}
{\it
В области параметров $\lambda\gs 0$, $0 \ls b \ls a$ перестройки $(S,H)$-диаграмм системы с двумя степенями свободы $\mam$ происходят тогда и только тогда, когда точка $(a, b,\lambda)$
пересекает разделяющее множество $\Theta$, состоящее из поверхностей

$\gamma_{1,2}: \quad \lambda=\sqrt{2}\left(\sqrt{a} \mp \sqrt{b}\right), \qquad b \in [0,a];$

$\gamma_3: \quad \lambda=\ds{\frac{a+b}{\sqrt{a-b}}}, \qquad b \in [0,a);$

$\gamma_4: \quad \lambda=\ds{\frac{a-b}{\sqrt{a+b}}}, \qquad b \in [0,a];$

$\gamma_5: \quad \lambda^4= {\ds{\frac{a^6-33 a^4 b^2-33 a^2 b^4+b^6+(a^4+14a^2b^2+b^4)^{3/2}}{54a^2b^2}}}, \quad b \in (0,1];$

$\gamma_6: \quad \lambda^2=\ds{\frac{(a-b)^2}{2\sqrt{a b}}},  \qquad b \in (0, a];$

$\gamma_7: \quad \lambda=\ds{\frac{a^2-b^2}{2^{3/4}(a^2+b^2)^{3/4}}}, \qquad b \in [0,a];$

$\gamma_8: \quad \lambda=\ds{\sqrt{2\,a} \, \frac{z^2-1}{z \sqrt{z}}}, \quad \ds{b=a \frac{(3-z^2)^{3/2}\sqrt{1+z^2}}{4 z^3}}, \qquad 1 \ls z \ls \sqrt{3}.$

\noindent Эти поверхности делят область параметров на $16$ подобластей со структурно устойчивыми диаграммами.
}
\end{theorem}

\begin{figure}[ht]
\centering
\includegraphics[width=0.8\textwidth, keepaspectratio]{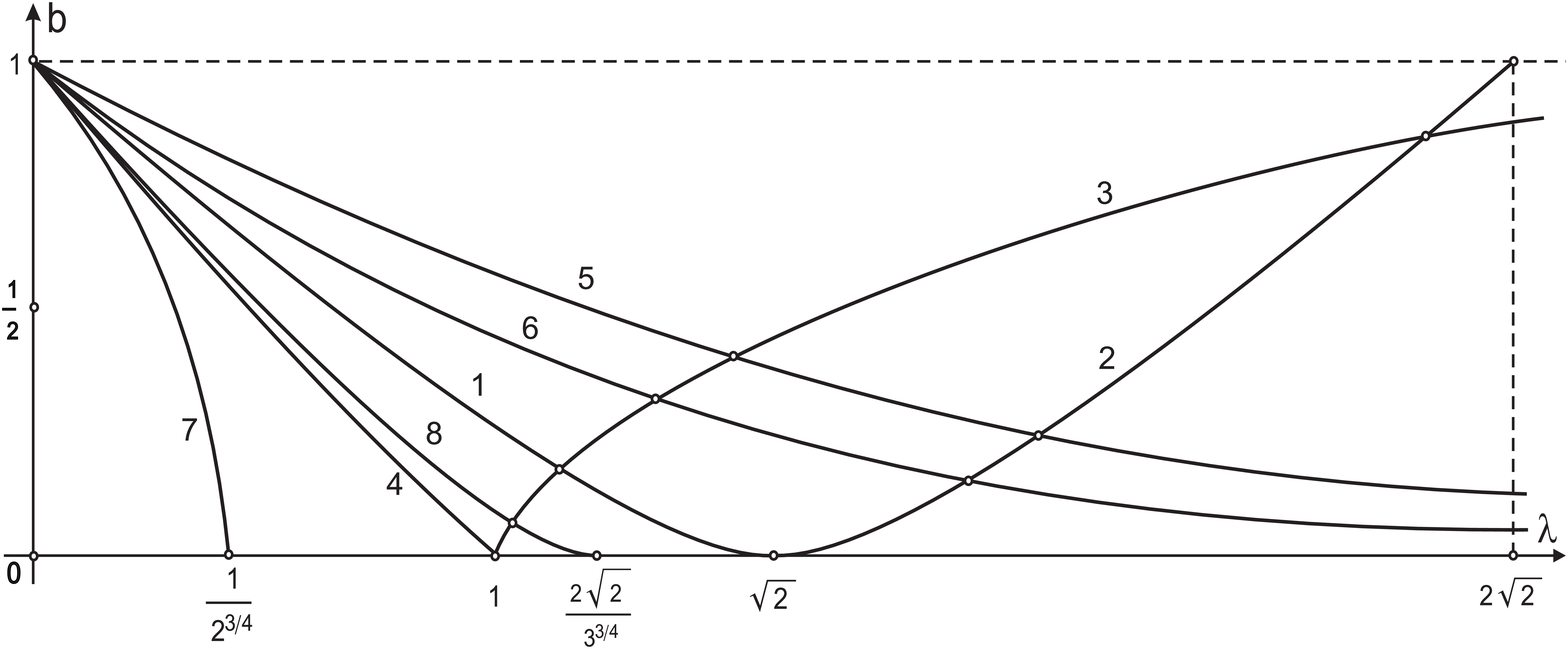}
\caption{Разделяющее множество для классификации диаграмм.}\label{fig:01}
\end{figure}

Поскольку, как отмечалось, всегда можно считать $a=1$, то будем говорить о разделяющих кривых $\gamma_i$ в полуполосе \eqref{eq:12} плоскости $\bR^2(\lambda,b)$. Они показаны на рис.~\ref{fig:01}. На каждой из них $\lambda$ есть однозначная функция от $b$ (явно выписанная на всех множествах, кроме $\gamma_8$, а на $\gamma_8$ имеем $\lambda^{\prime}_z>0$ при $z>0$). Обозначим такую функцию на $\gamma_i$ через $\lambda_i(b)$ ($i=1,\ldots,8$).

\begin{figure}[ht]
\makeatletter
\renewcommand\thesubfigure{\alph{subfigure}}
\renewcommand\p@subfigure{}
\makeatother
\centering
      \begin{subfigure}[b]{0.45\textwidth}
                \centering
                \includegraphics[width=\textwidth]{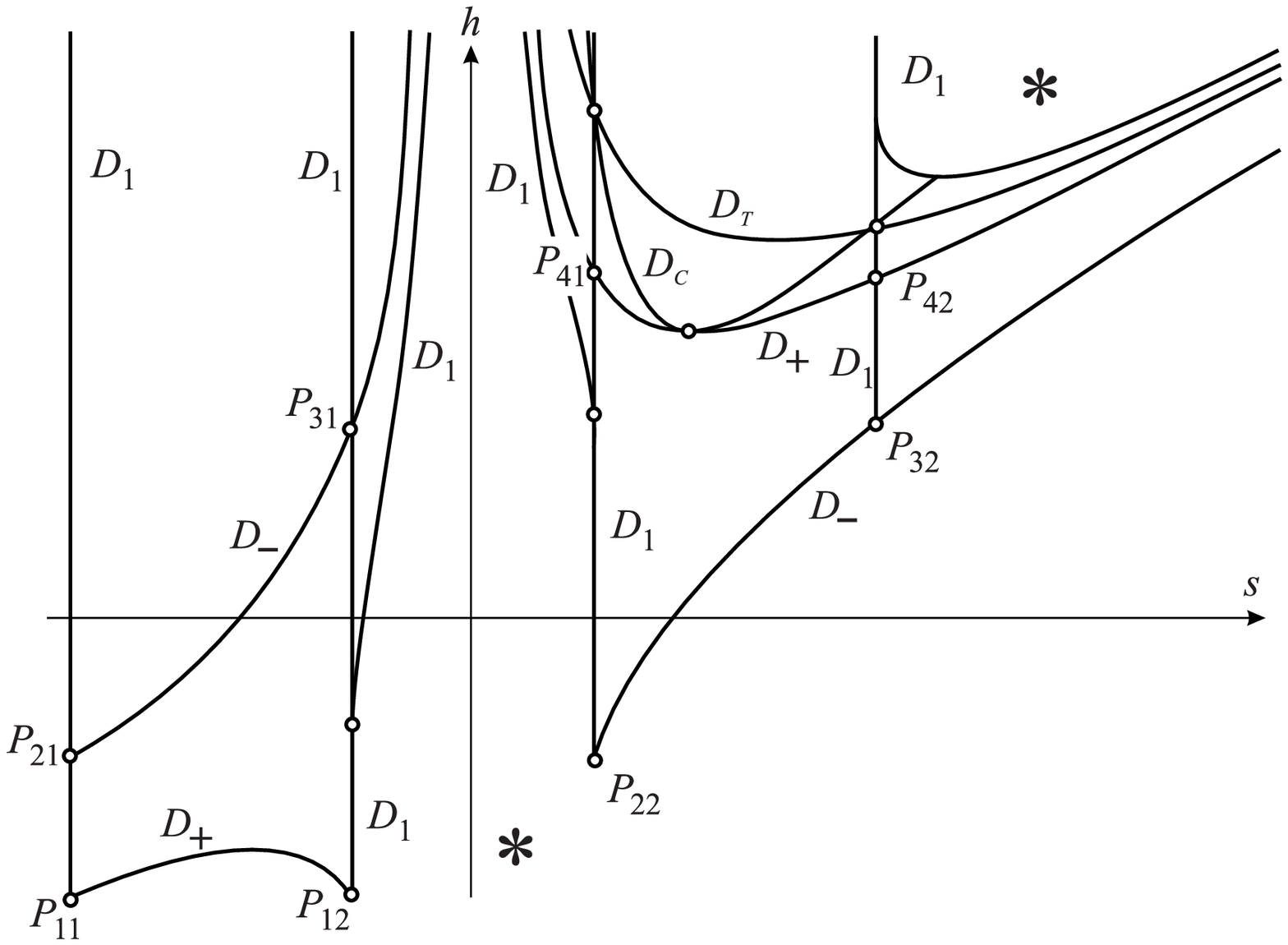}
                \caption{$b=0.3,\,\lambda=0$}
                \label{fig:bif0}
        \end{subfigure}%
        \begin{subfigure}[b]{0.45\textwidth}
                \centering
                \includegraphics[width=\textwidth]{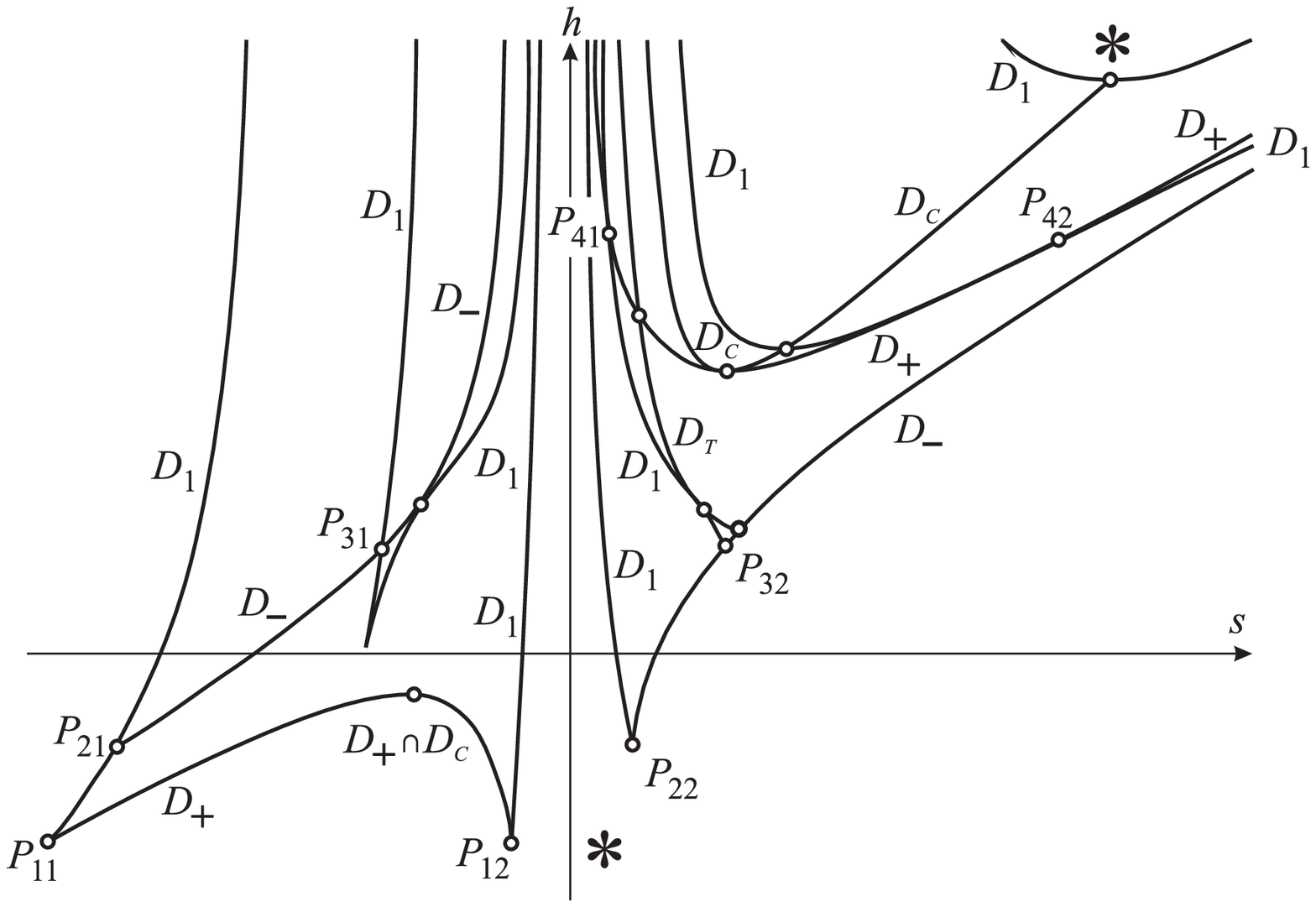}
                \caption{$b=0.3,\,\lambda =1.2$}
                \label{fig:bif1}
        \end{subfigure}
         \caption{Примеры $(S,H)$-диаграмм.}\label{fig:02}
\end{figure}

На рис.~\ref{fig:02} показана известная $(S,H)$-диаграмма при $\lambda=0$ \cite{KhShRCD06} и одна из возможных диаграмм при $\lambda>0$. Видно, что структура диаграммы в общем случае достаточно сложна. Звездочкой помечены компоненты дополнения к диаграмме, в которых интегральные многообразия пусты.


\makeatletter
\renewcommand\thesubfigure{\arabic{subfigure}}
\renewcommand\p@subfigure{}
\makeatother
\def\coff{0.25}
\def\spa{\hspace{2cm}}
\def\mcof{0.32}

\clearpage

\begin{figure}[hp]
        \centering
\subcaptionbox{$\lambda < \lambda_1$\label{fig:03_01}}
[\mcof\linewidth]{\includegraphics[width=\coff\textwidth]{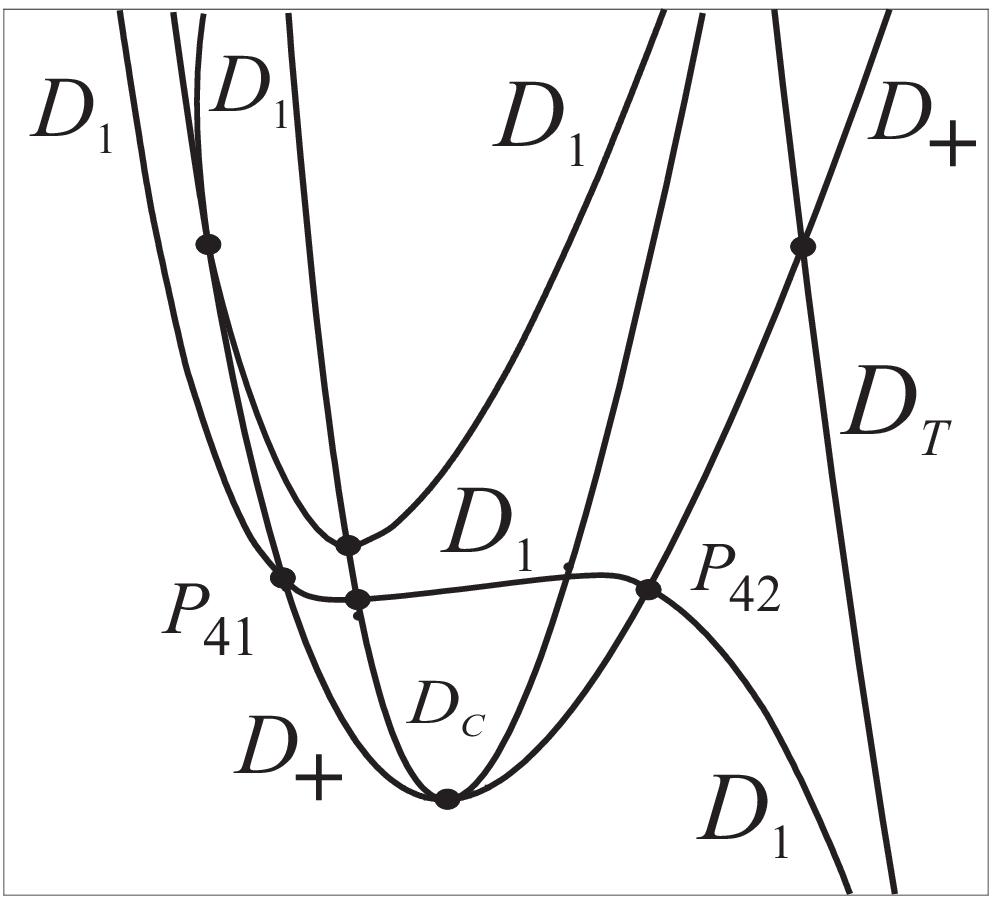}}%
\subcaptionbox{$\lambda = \lambda_1$\label{fig:03_02}}%
[\mcof\linewidth]{\includegraphics[width=\coff\textwidth]{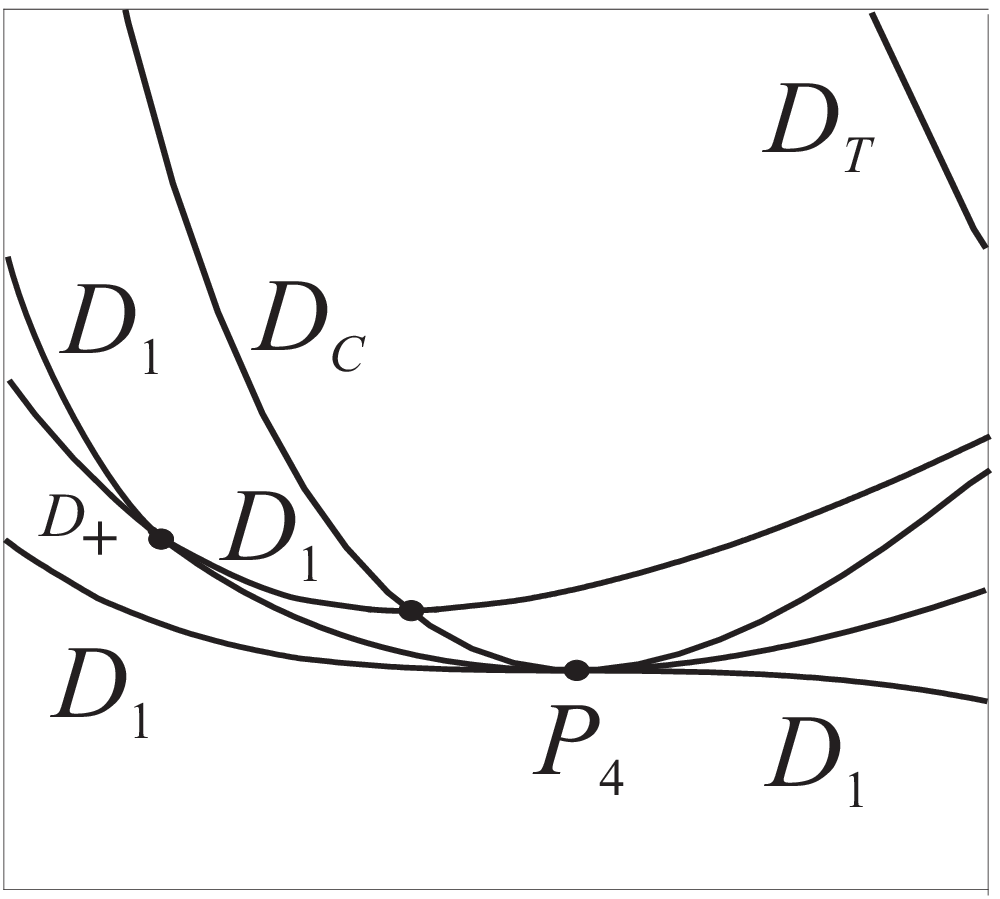}}
\subcaptionbox{$\lambda > \lambda_1$\label{fig:03_03}}%
[\mcof\linewidth]{\includegraphics[width=\coff\textwidth]{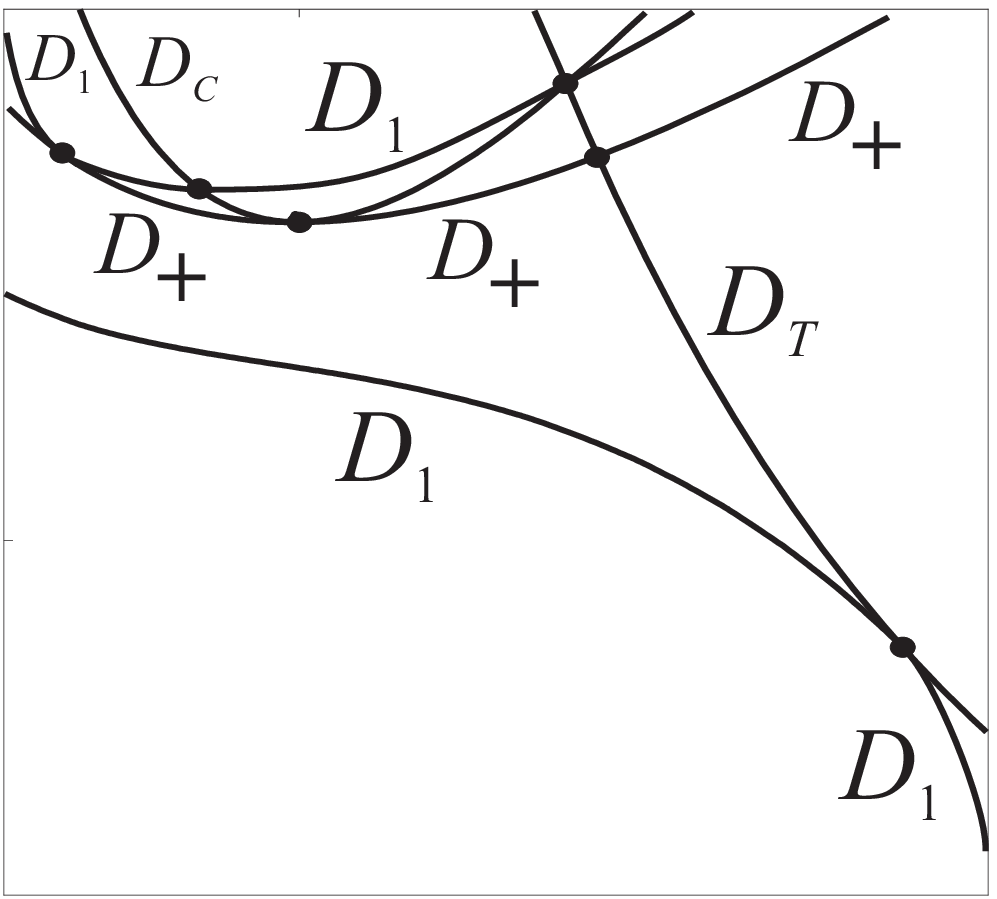}}

\subcaptionbox{$\lambda < \lambda_2$\label{fig:03_04}}
[\mcof\linewidth]{\includegraphics[width=\coff\textwidth]{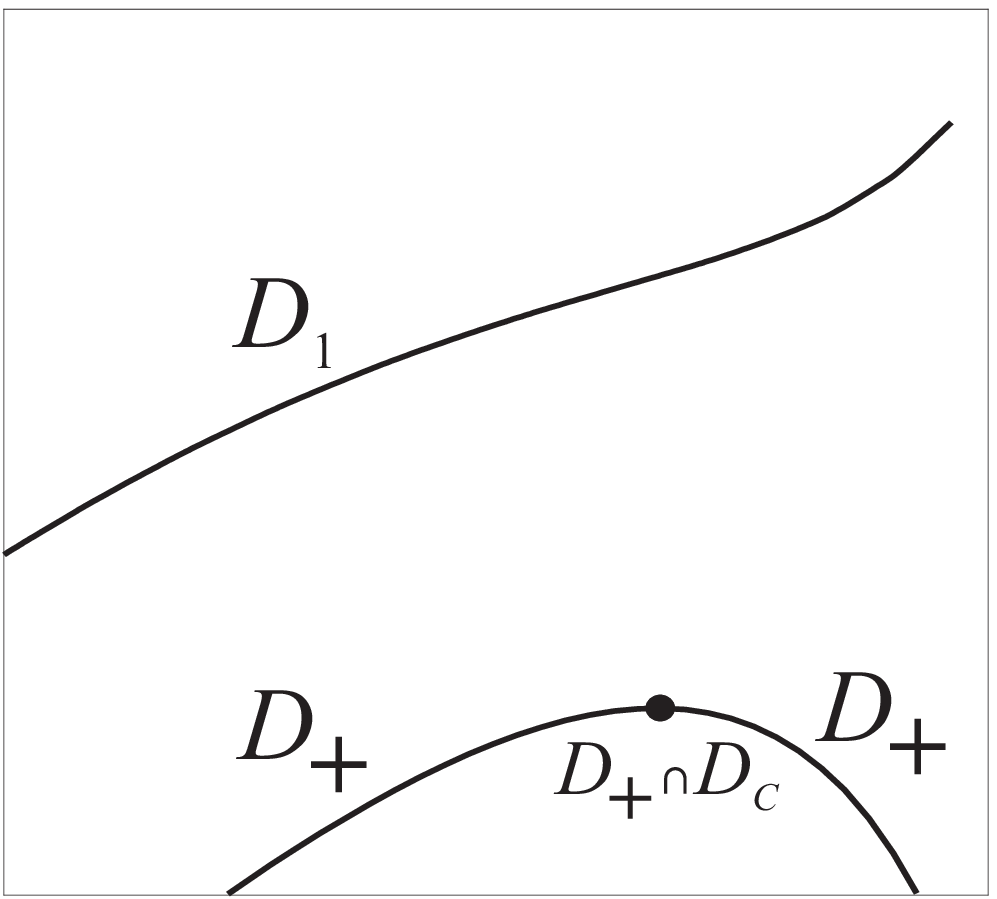}}%
\subcaptionbox{$\lambda = \lambda_2$\label{fig:03_05}}%
[\mcof\linewidth]{\includegraphics[width=\coff\textwidth]{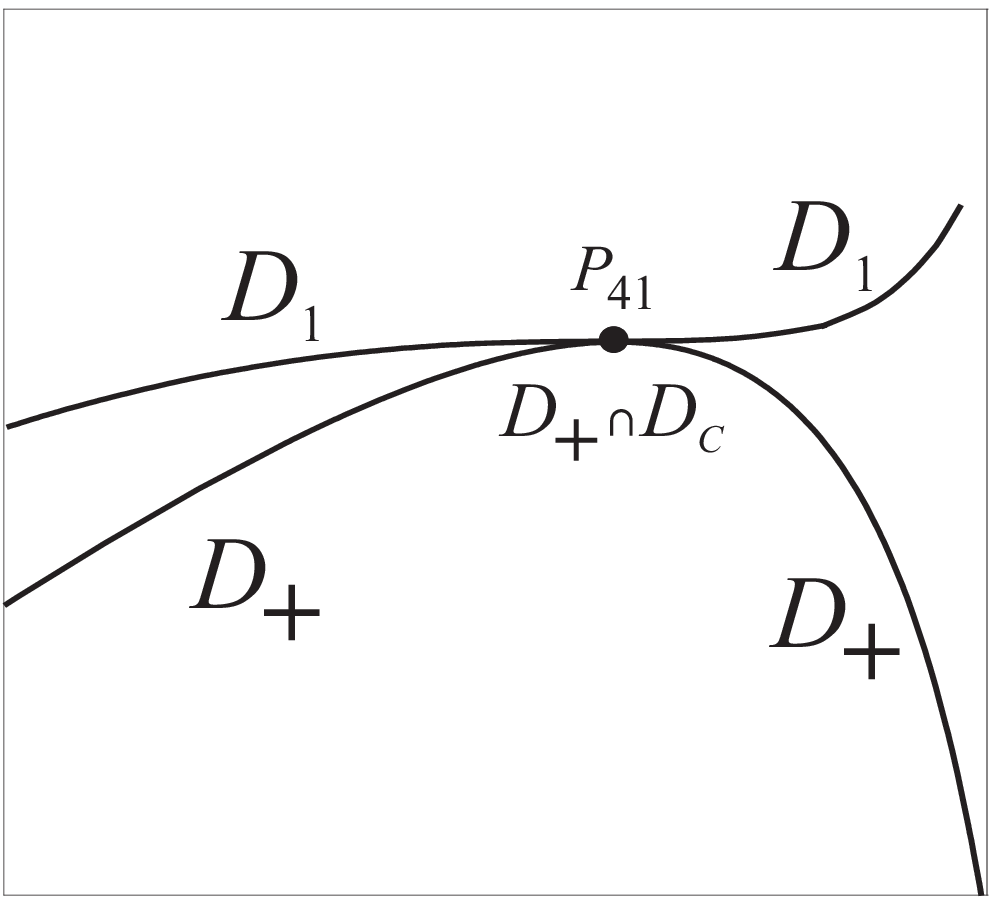}}
\subcaptionbox{$\lambda > \lambda_2$\label{fig:03_06}}%
[\mcof\linewidth]{\includegraphics[width=\coff\textwidth]{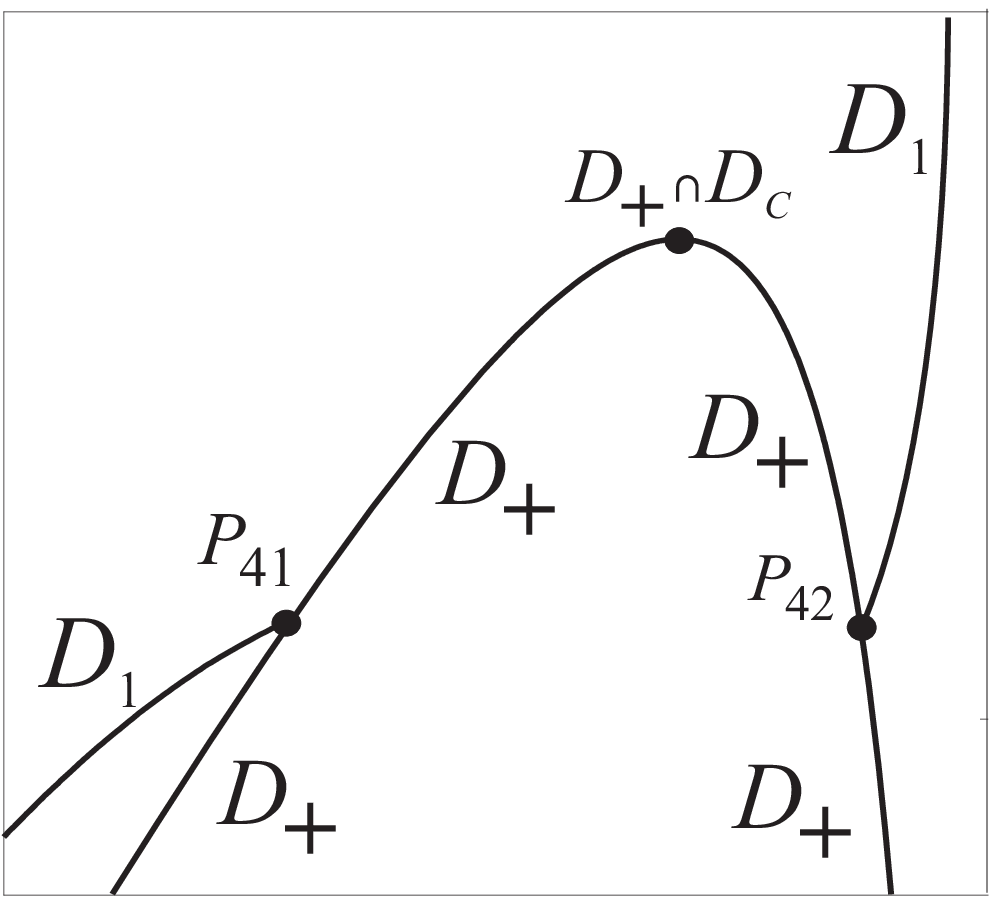}}

\subcaptionbox{$\lambda < \lambda_3$\label{fig:03_07}}
[\mcof\linewidth]{\includegraphics[width=\coff\textwidth]{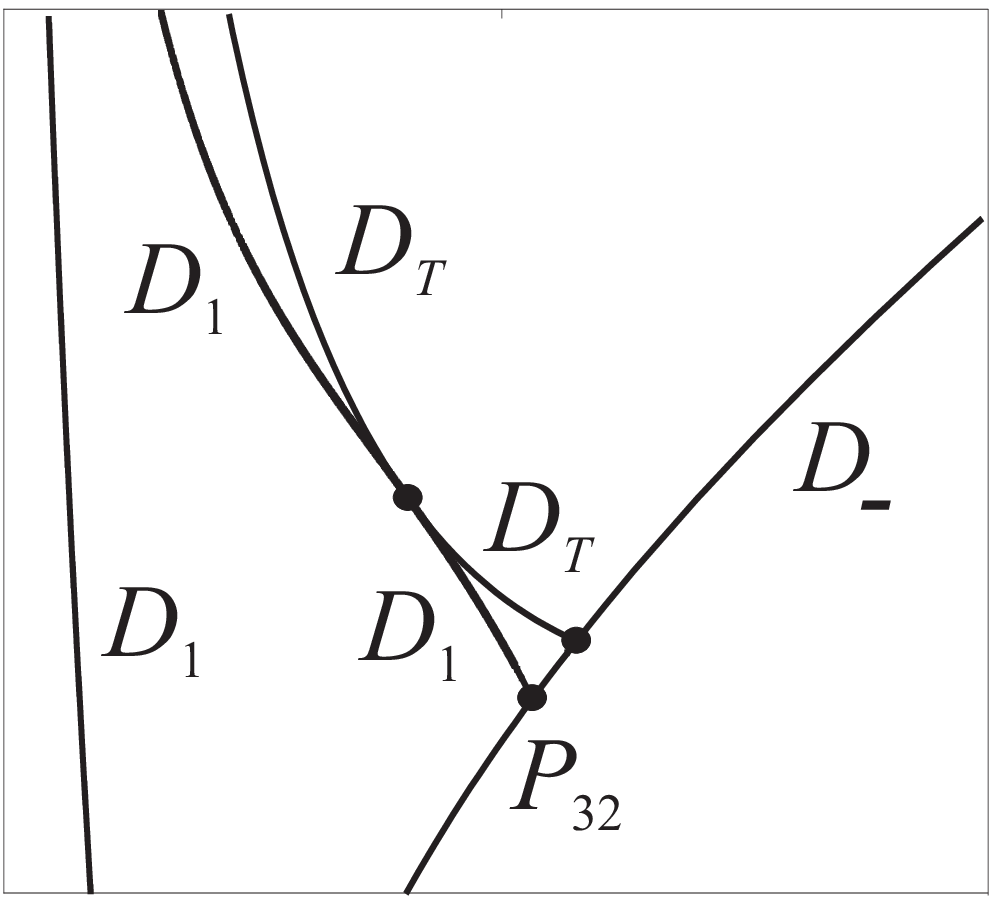}}%
\subcaptionbox{$\lambda = \lambda_3$\label{fig:03_08}}%
[\mcof\linewidth]{\includegraphics[width=\coff\textwidth]{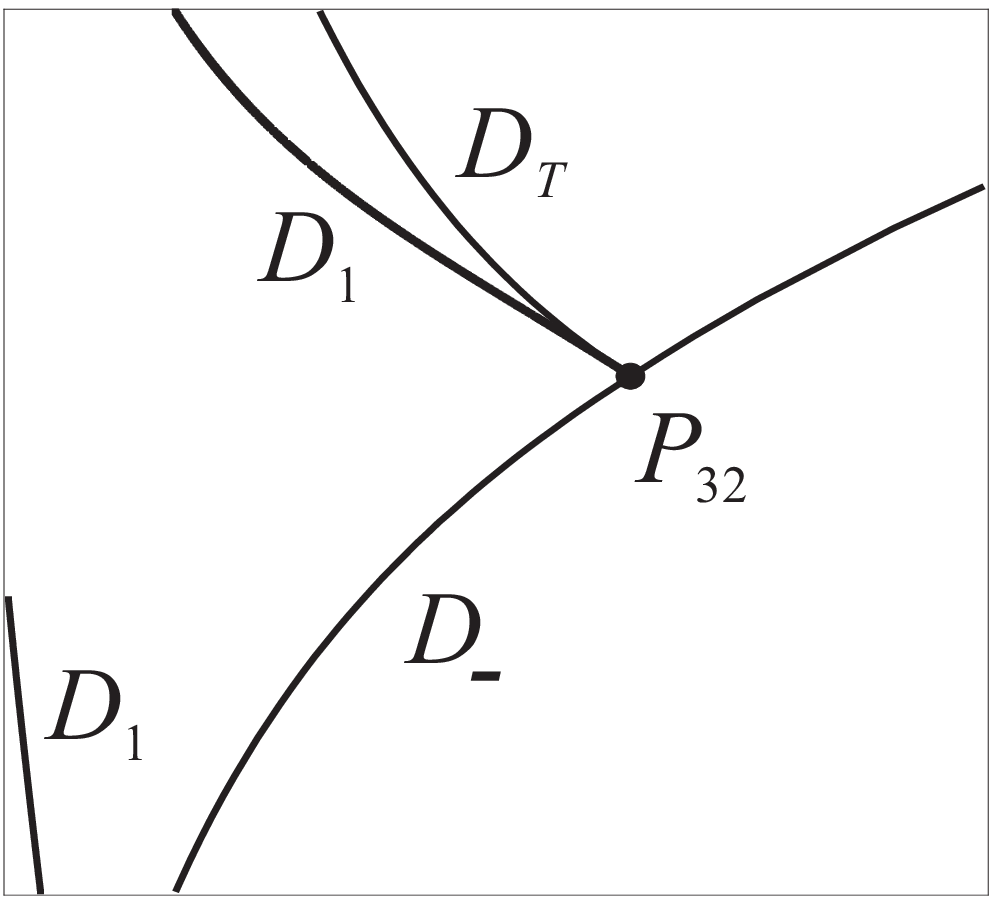}}
\subcaptionbox{$\lambda > \lambda_3$\label{fig:03_09}}%
[\mcof\linewidth]{\includegraphics[width=\coff\textwidth]{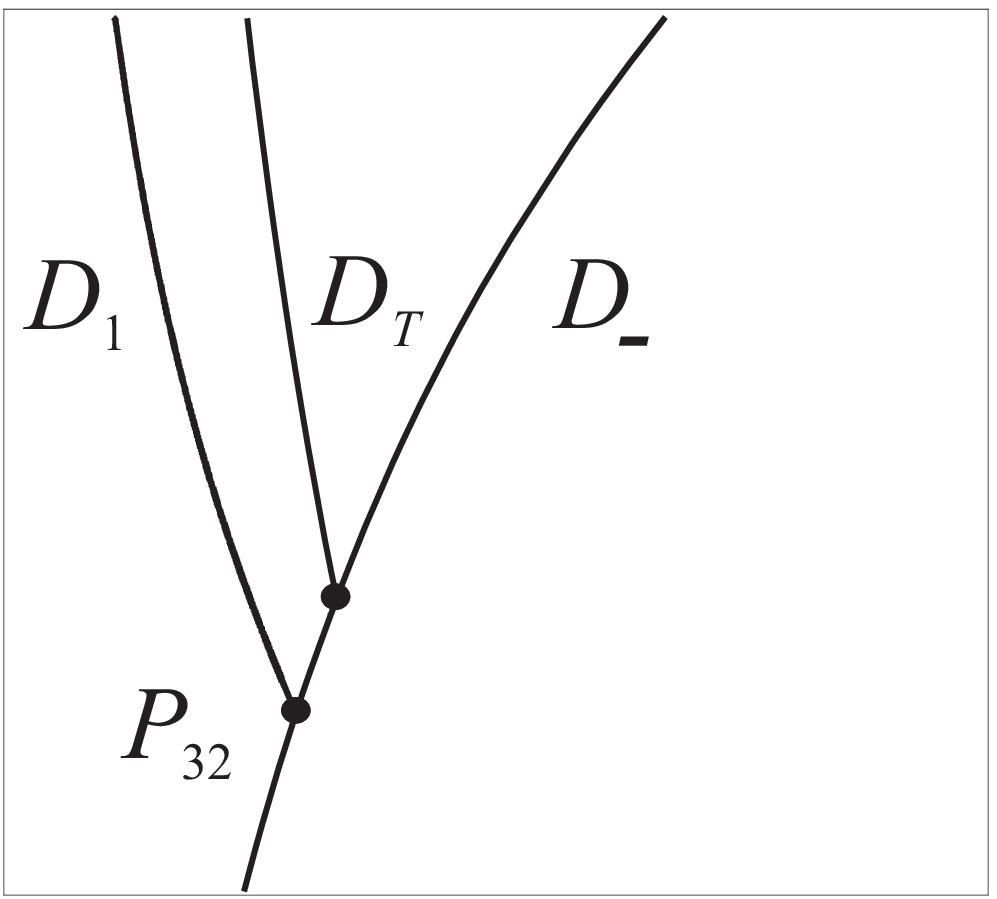}}

\subcaptionbox{$\lambda < \lambda_3$\label{fig:03_10}}
[\mcof\linewidth]{\includegraphics[width=\coff\textwidth]{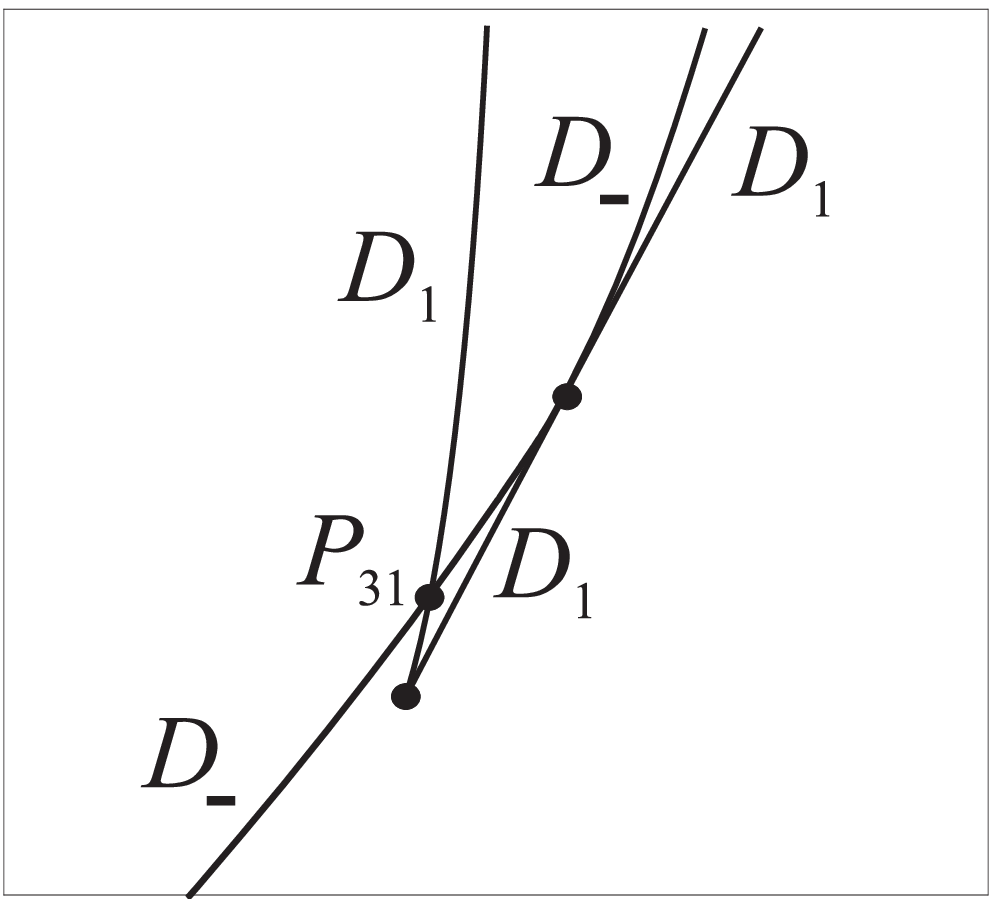}}%
\subcaptionbox{$\lambda = \lambda_3$\label{fig:03_11}}%
[\mcof\linewidth]{\includegraphics[width=\coff\textwidth]{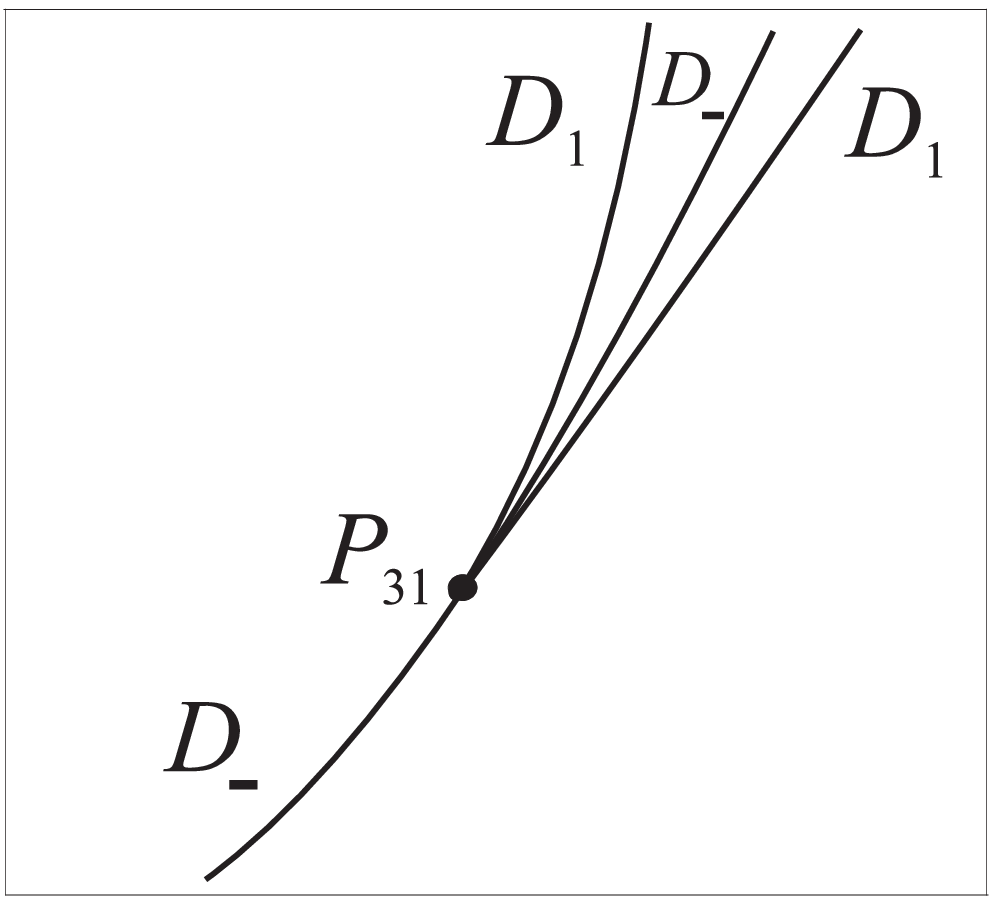}}
\subcaptionbox{$\lambda > \lambda_3$\label{fig:03_12}}%
[\mcof\linewidth]{\includegraphics[width=\coff\textwidth]{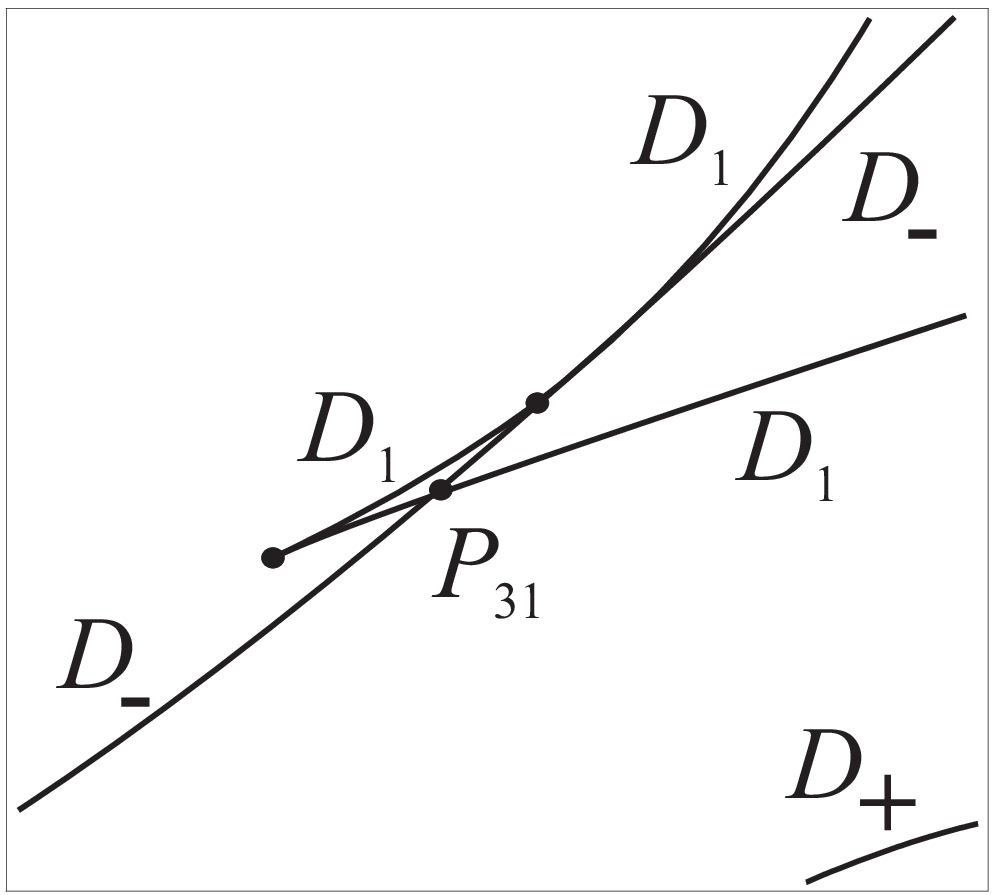}}

\subcaptionbox{$\lambda < \lambda_4$\label{fig:03_13}}
[\mcof\linewidth]{\includegraphics[width=\coff\textwidth]{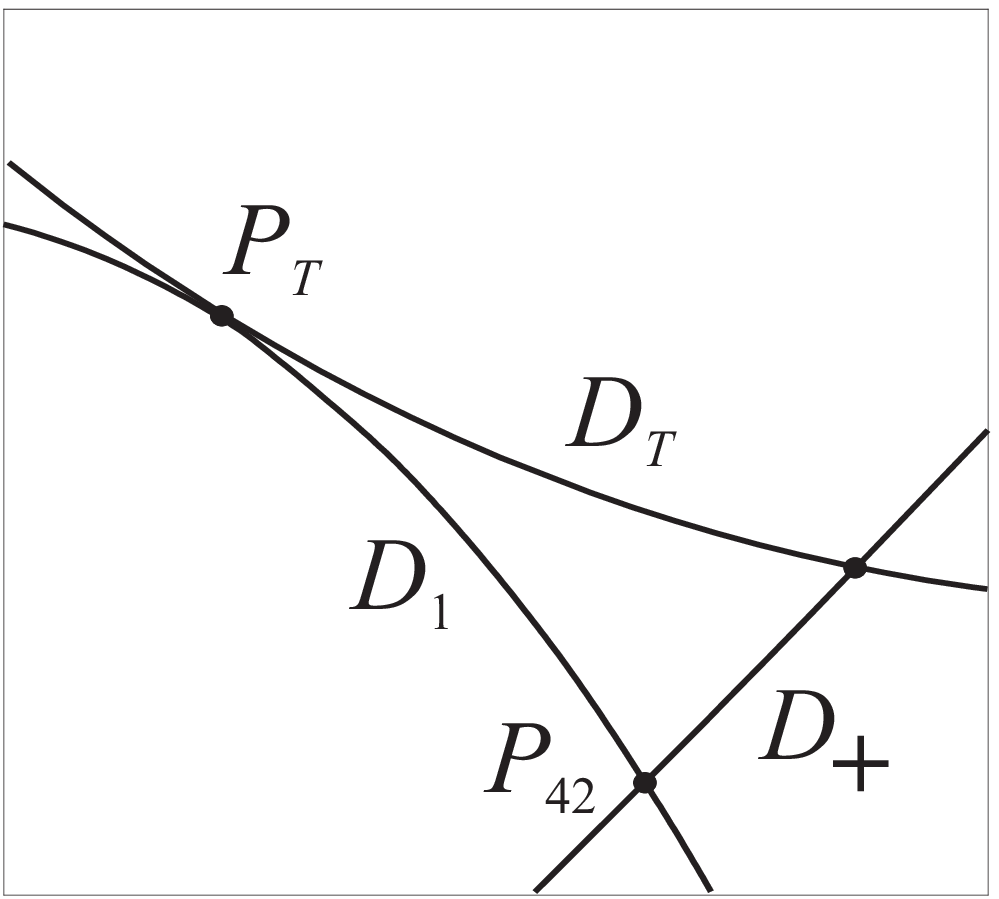}}%
\subcaptionbox{$\lambda = \lambda_4$\label{fig:03_14}}%
[\mcof\linewidth]{\includegraphics[width=\coff\textwidth]{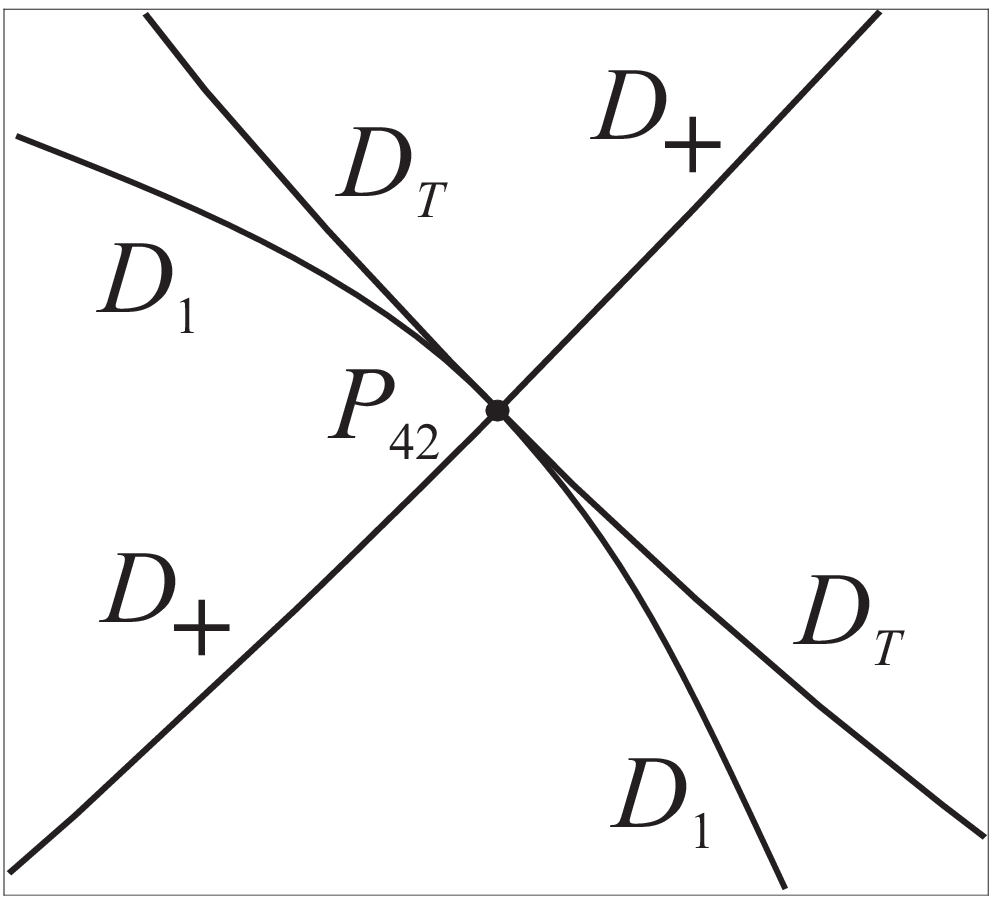}}
\subcaptionbox{$\lambda > \lambda_4$\label{fig:03_15}}%
[\mcof\linewidth]{\includegraphics[width=\coff\textwidth]{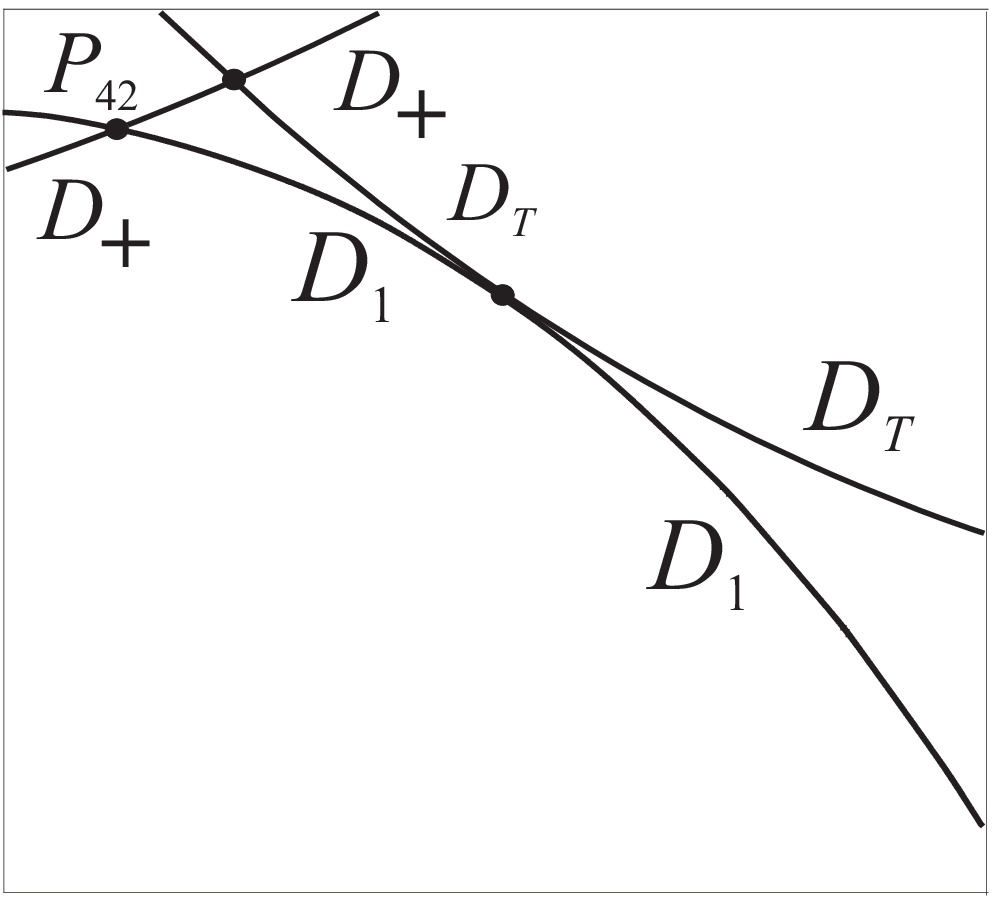}}
\end{figure}

\clearpage

\begin{figure}[h!]
        \centering
\subcaptionbox{$\lambda < \lambda_5$\label{fig:03_16}}
[\mcof\linewidth]{\includegraphics[width=\coff\textwidth]{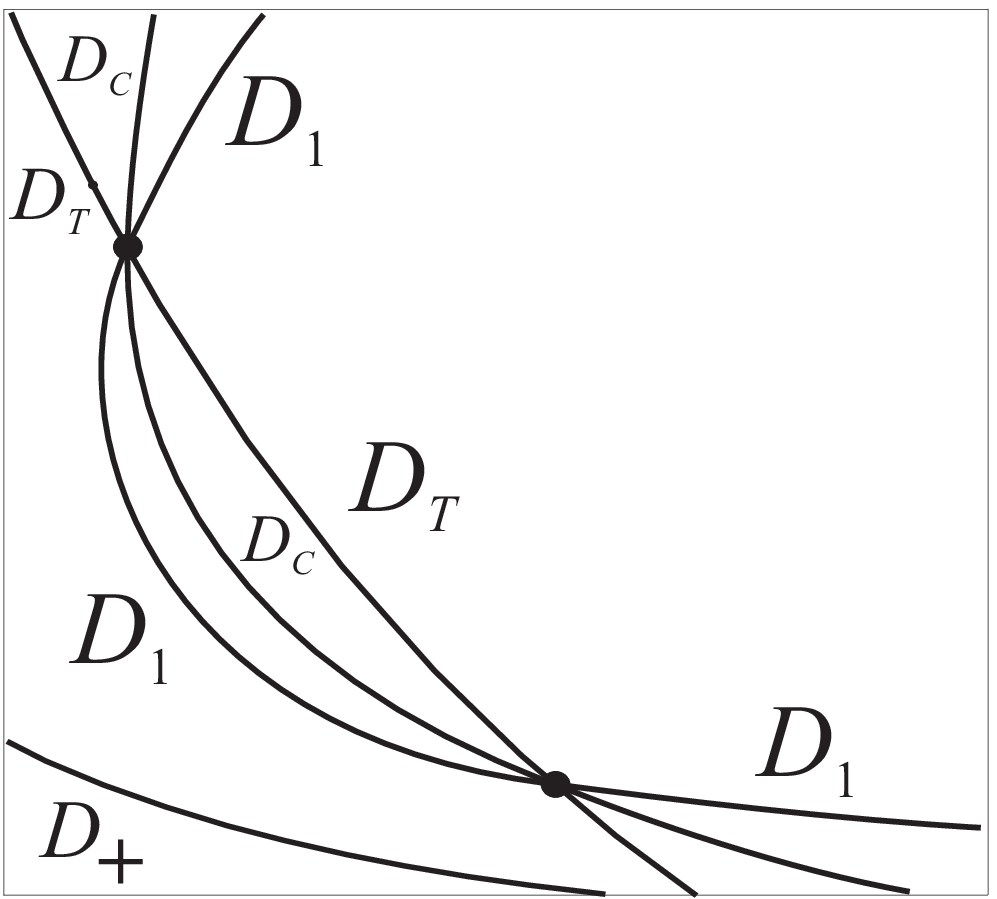}}%
\subcaptionbox{$\lambda = \lambda_5$\label{fig:03_17}}%
[\mcof\linewidth]{\includegraphics[width=\coff\textwidth]{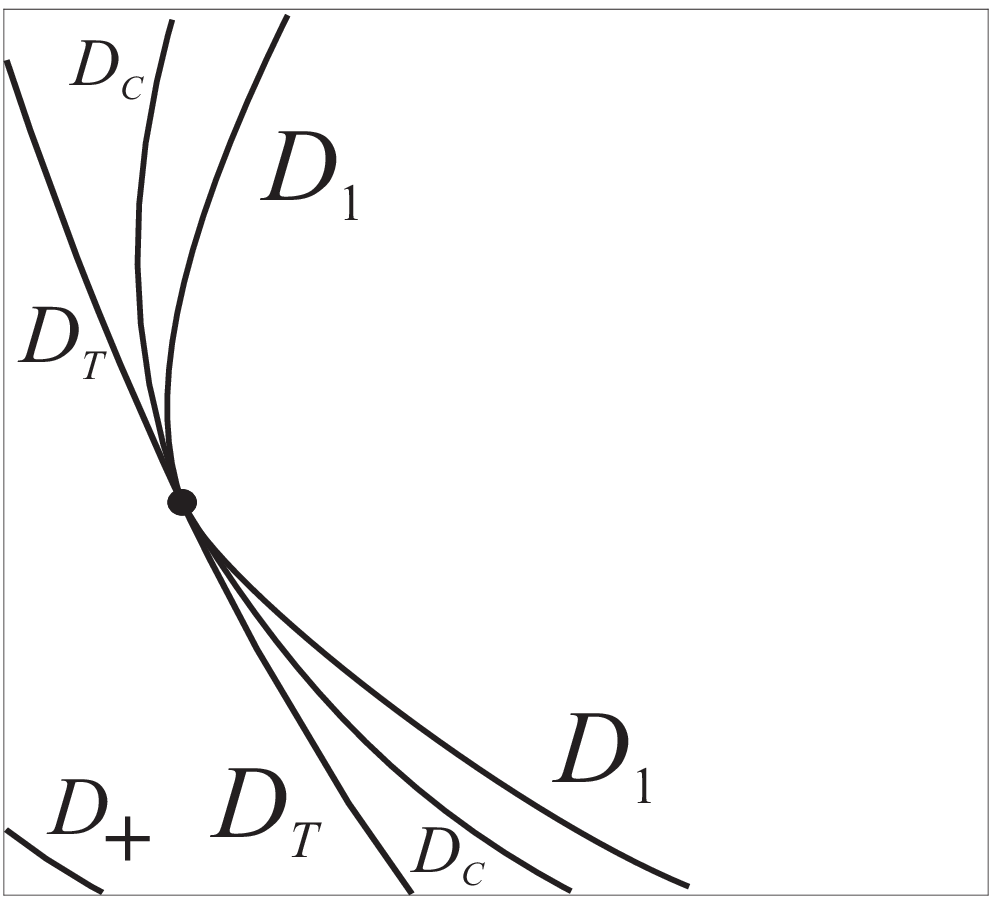}}
\subcaptionbox{$\lambda > \lambda_5$\label{fig:03_18}}%
[\mcof\linewidth]{\includegraphics[width=\coff\textwidth]{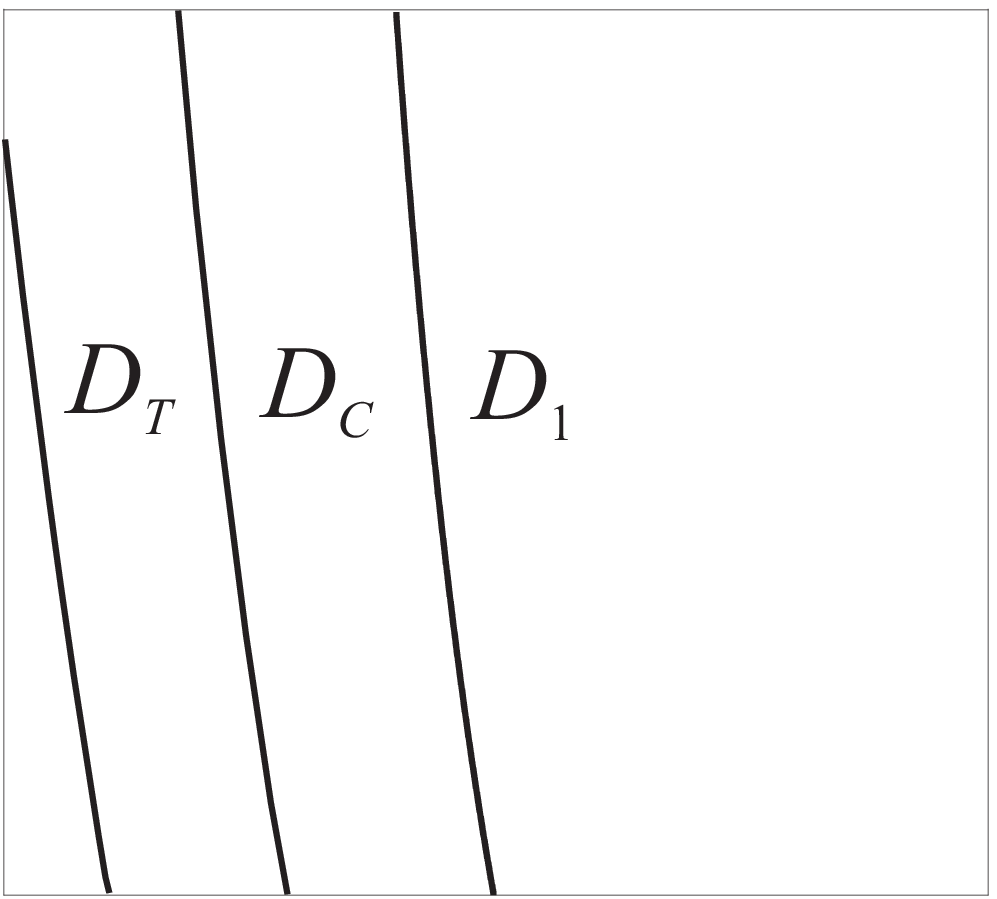}}

\subcaptionbox{$\lambda < \lambda_6$\label{fig:03_19}}
[\mcof\linewidth]{\includegraphics[width=\coff\textwidth]{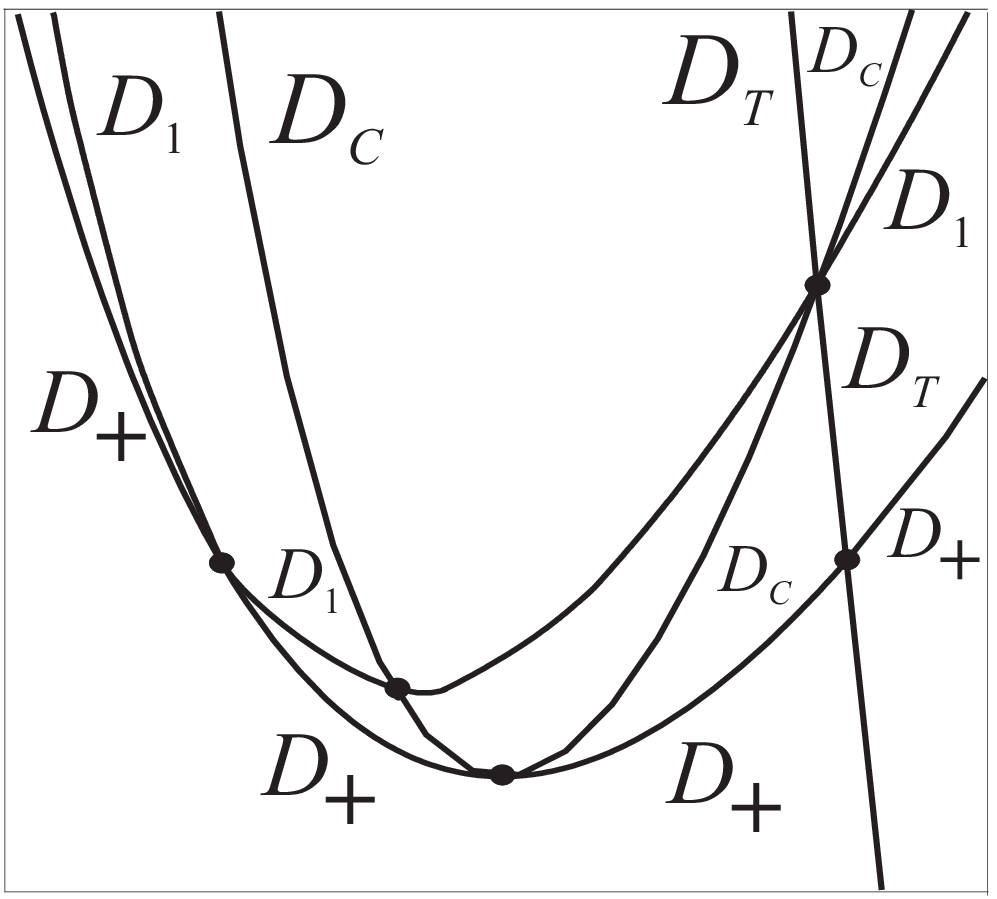}}%
\subcaptionbox{$\lambda = \lambda_6$\label{fig:03_20}}%
[\mcof\linewidth]{\includegraphics[width=\coff\textwidth]{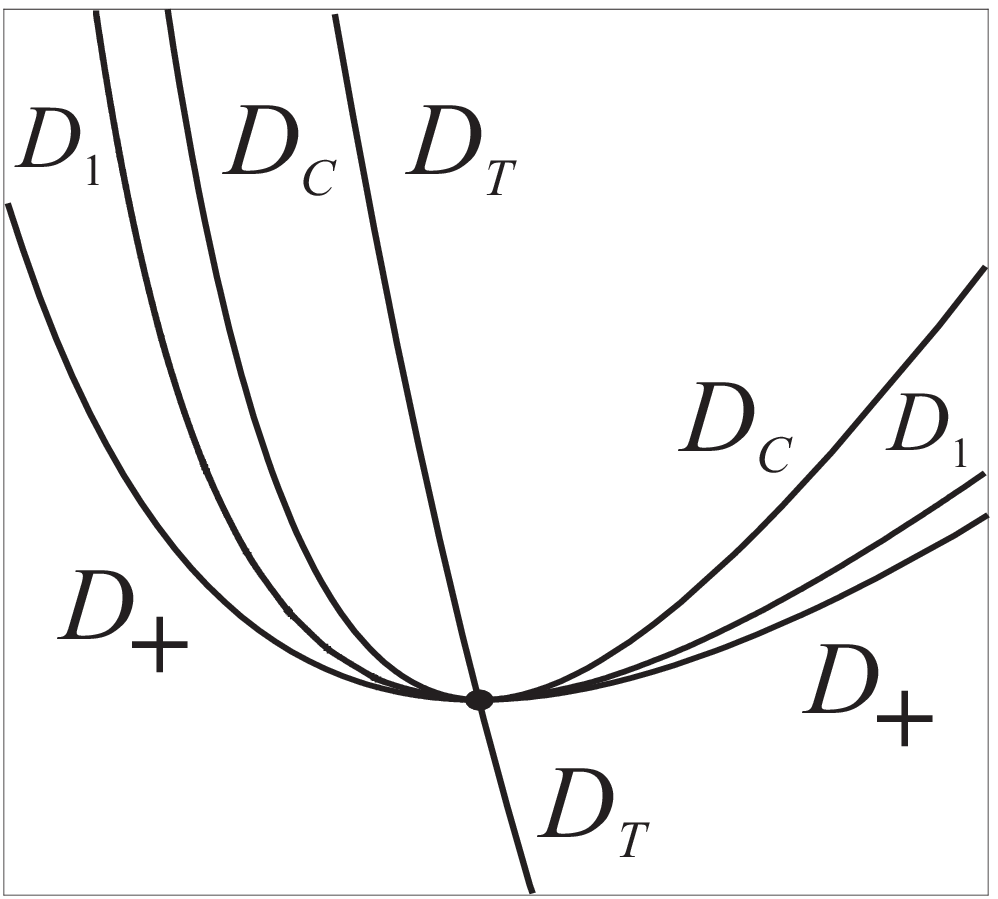}}
\subcaptionbox{$\lambda > \lambda_6$\label{fig:03_21}}%
[\mcof\linewidth]{\includegraphics[width=\coff\textwidth]{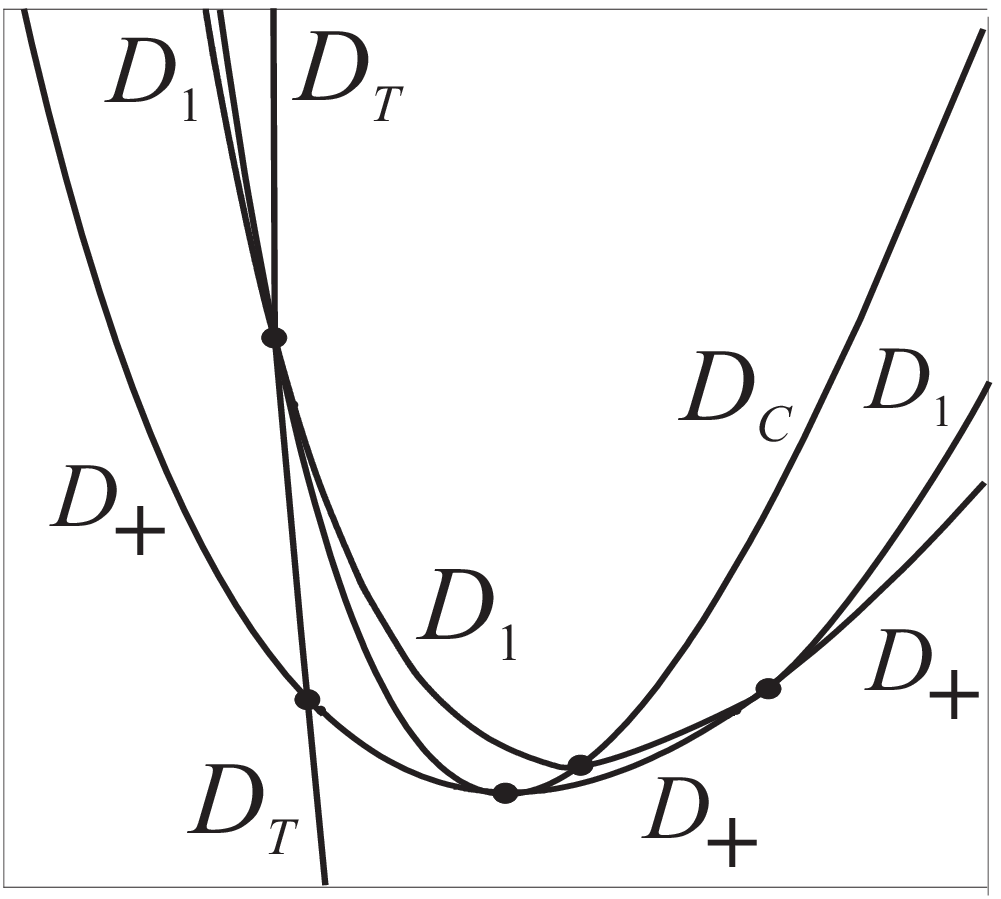}}

\subcaptionbox{$\lambda < \lambda_7$\label{fig:03_22}}
[\mcof\linewidth]{\includegraphics[width=\coff\textwidth]{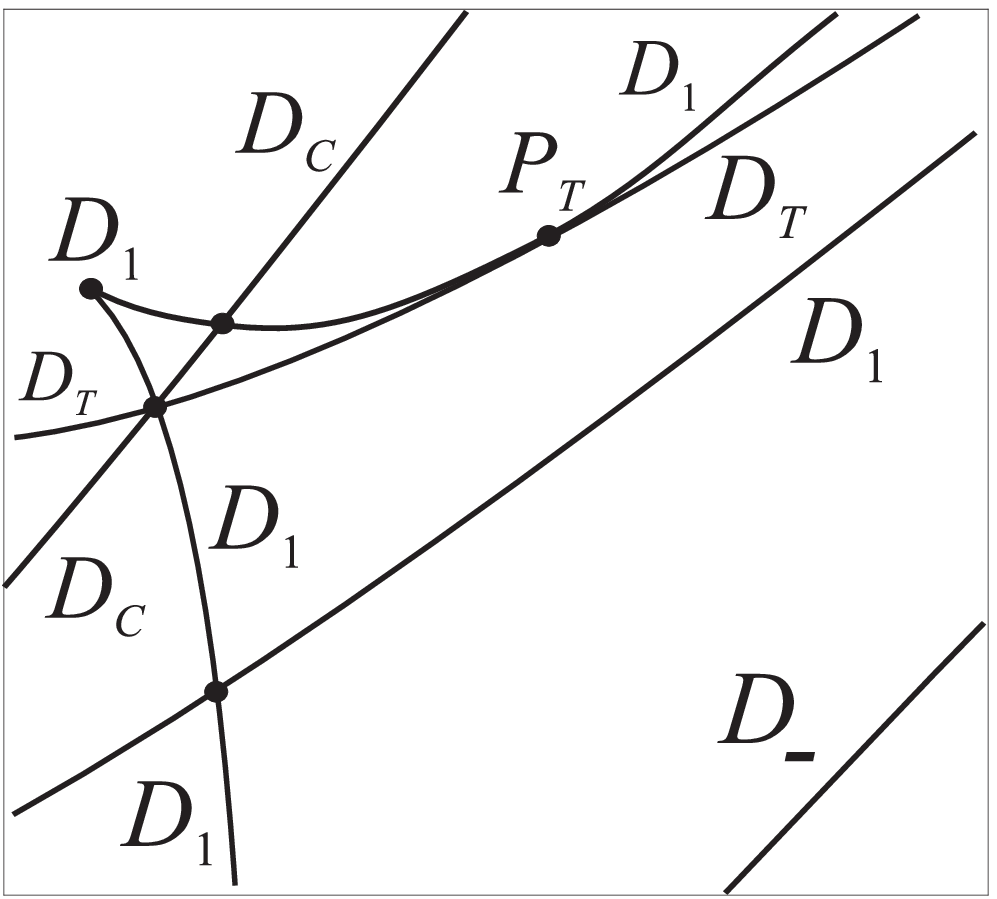}}%
\subcaptionbox{$\lambda = \lambda_7$\label{fig:03_23}}%
[\mcof\linewidth]{\includegraphics[width=\coff\textwidth]{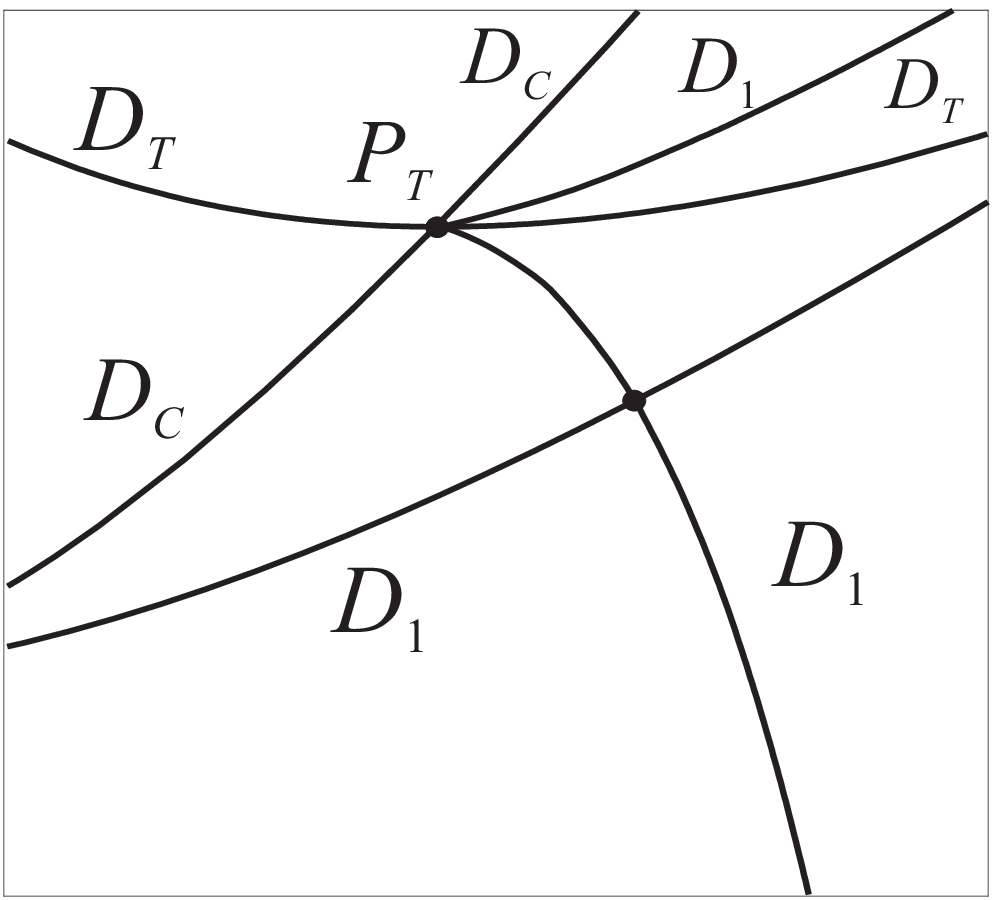}}
\subcaptionbox{$\lambda > \lambda_7$\label{fig:03_24}}%
[\mcof\linewidth]{\includegraphics[width=\coff\textwidth]{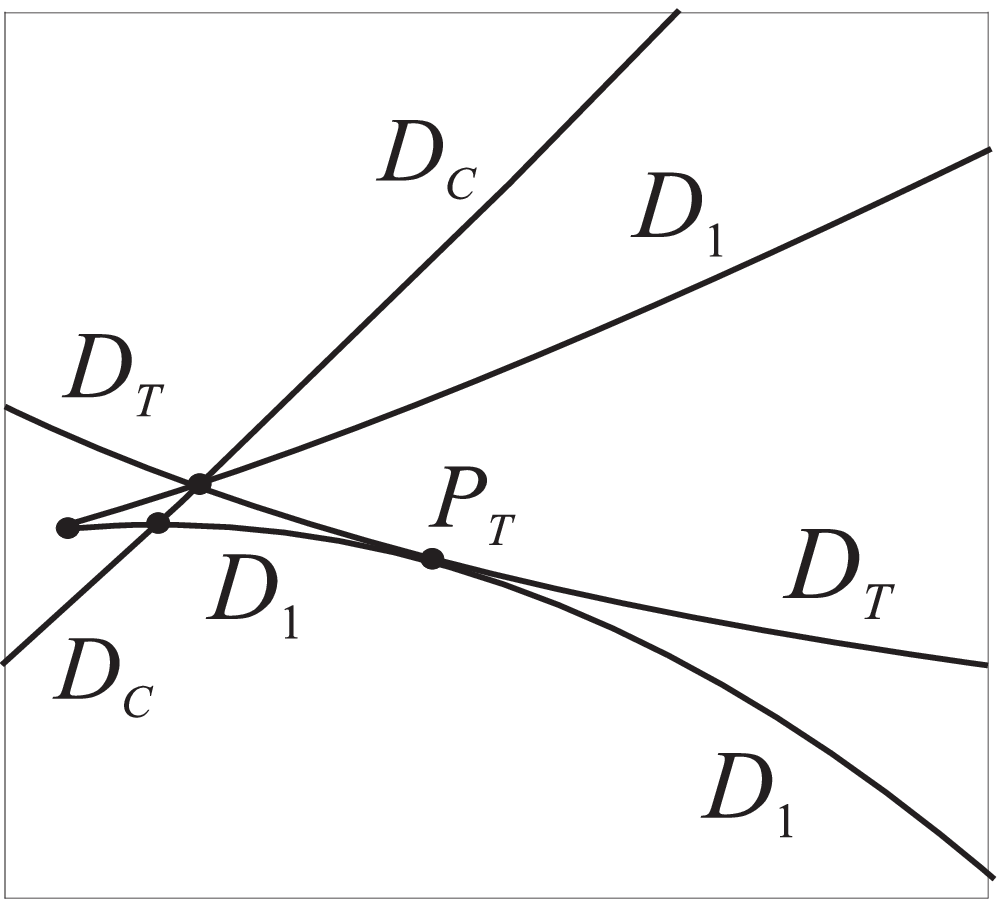}}

\subcaptionbox{$\lambda < \lambda_8$\label{fig:03_25}}
[\mcof\linewidth]{\includegraphics[width=\coff\textwidth]{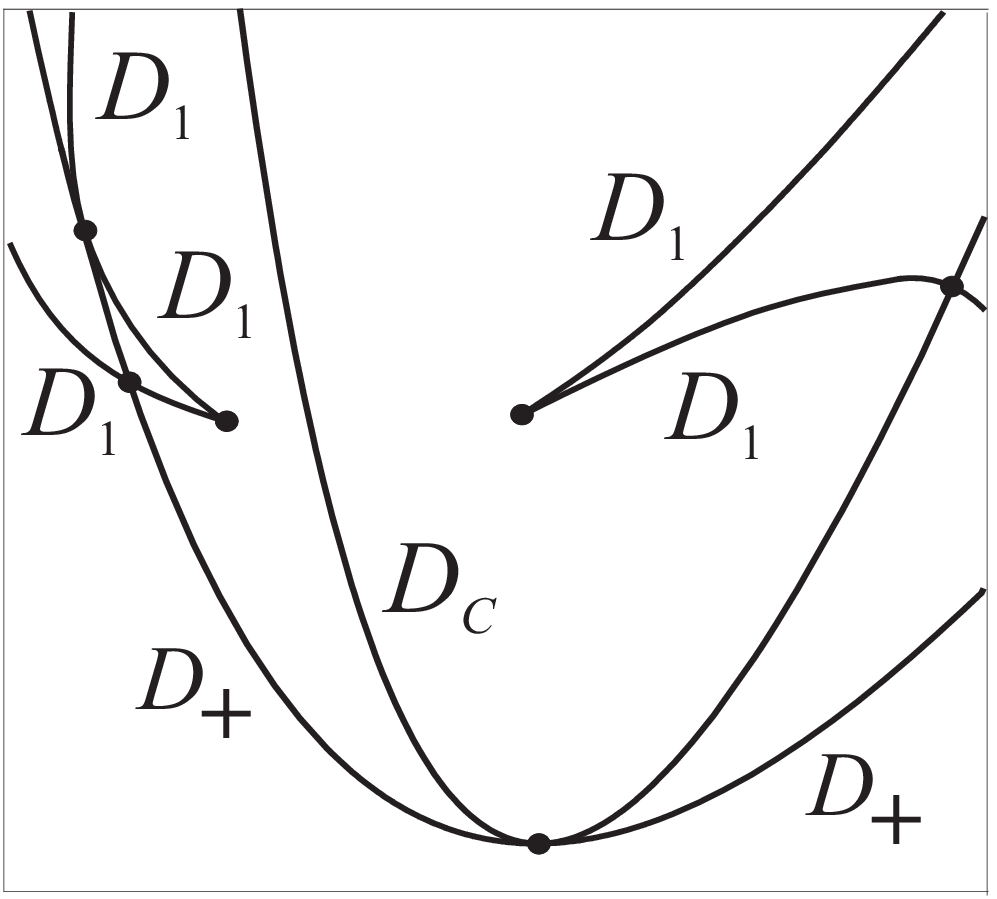}}%
\subcaptionbox{$\lambda = \lambda_8$\label{fig:03_26}}%
[\mcof\linewidth]{\includegraphics[width=\coff\textwidth]{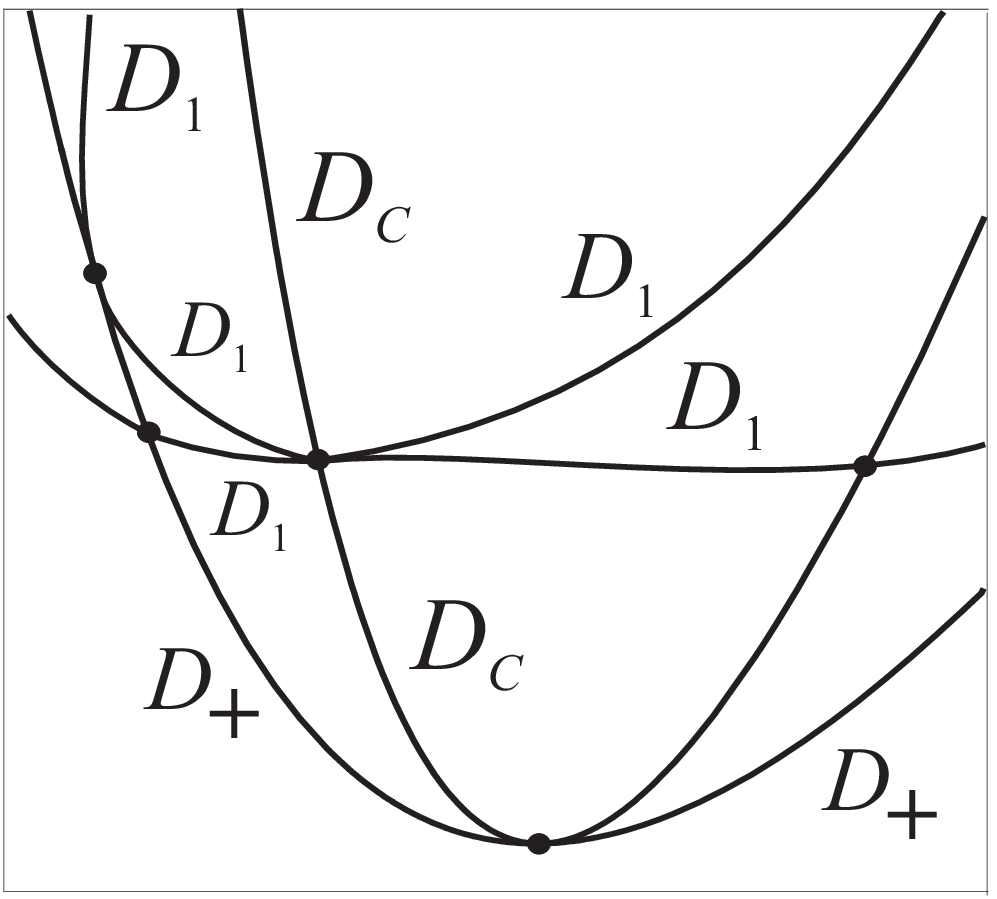}}
\subcaptionbox{$\lambda > \lambda_8$\label{fig:03_27}}%
[\mcof\linewidth]{\includegraphics[width=\coff\textwidth]{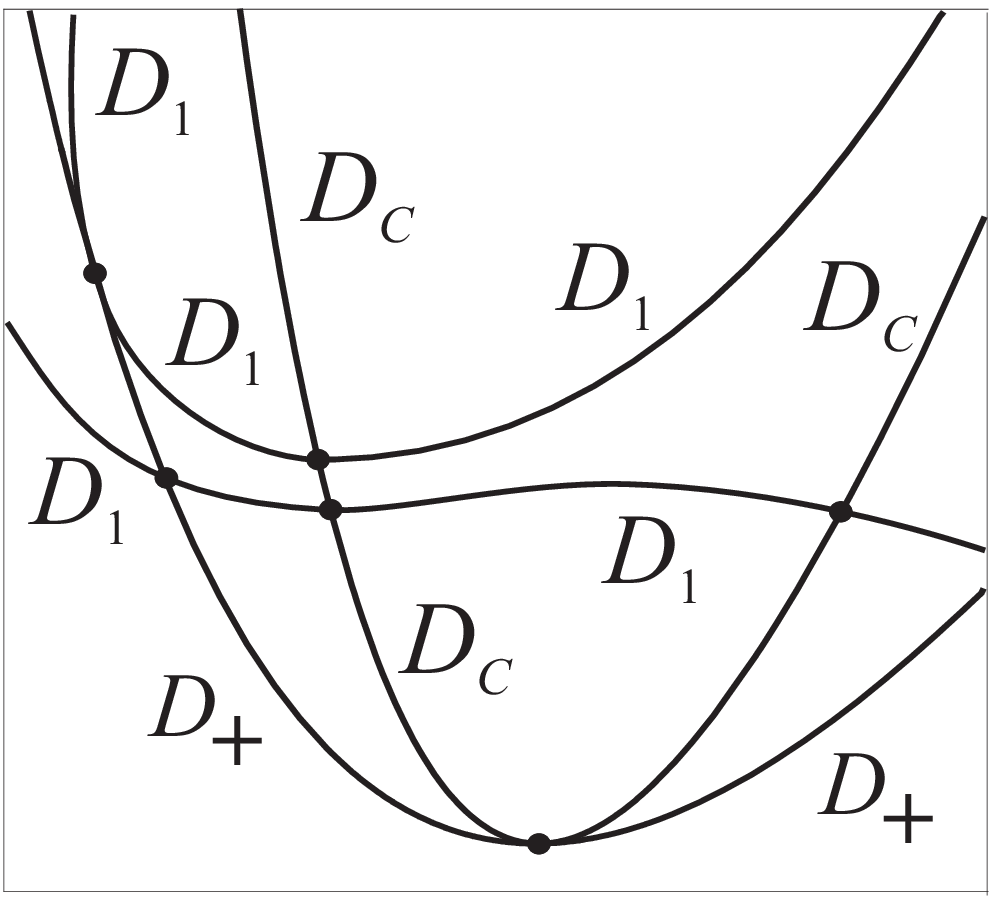}}
        \caption{Перестройки в диаграммах.}\label{fig:03}
\end{figure}


Для того, чтобы получить возможность построить и проанализировать любую диаграмму, на рис.~\ref{fig:03} приведены перестройки фрагментов диаграммы при пересечении разделяющих кривых. Подрисунки на рис.~\ref{fig:03} организованы в тройки, указывающие фрагменты слева от соответствующей кривой, на самой кривой (имеется кратная точка среди узловых) и справа от кривой. Для кривой $\gamma_3$ таких троек две, так как перестройки фрагментов происходят одновременно на удаленных друг от друга участках в окрестности точек $P_{31}$ и $P_{32}$. Таким образом, получены необходимые и достаточные условия бифуркаций в семействе диаграмм системы $\mam$ и явно указаны соответствующие перестройки, происходящие в окрестностях узловых точек. Следующий шаг~-- топологический анализ системы~-- будет выполнен в дальнейшем с учетом вычисленных ранее типов критических точек.

\begin{abstract}[en]
\title{The atlas of the diagrams for the generalization of the 4th Appelrot class of especially remarkable motions to a gyrostat in a double force field}
\author{P.E.\,Ryabov, G.E.\,Smirnov, M.P.\,Kharlamov}
For the system with two degrees of freedom, which is an analogue of the 4th Appelrot class for a gyrostat of the Kowalevski type in a double force field the problem of the classification of bifurcation diagrams is solved. The separating set is built and its completeness is proved. All transformations taking place in the diagrams are shown.
The results serve as a necessary part of solving the problem of obtaining the topological invariants for the Reyman\,--\,Semenov-Tian-Shansky system.
\end{abstract}

\begin{abstract}[uk]
\title{Атлас діаграм узагальнення 4-го класу особливо чудових рухів\\Аппельрота на гіростат в подвійному полі}
\author{П.Э.\,Рябов, Г.Э.\,Смирнов, М.П.\,Харламов}
У системі з двома мірами свободи, яка є аналогом 4-го класу Аппельрота для гіростата з умовами типа Ковалевською в подвійному полі, дана класифікація біфуркацiйних діаграм. Побудована розділяюча безліч і даний доказ його повноти. Представлені всі перетворення, що мають місце в діаграмах. Результати є необхідним кроком у вирішенні проблеми побудови топологічних інваріантів для інтегровано\"{\i} системи Реймана--Семенова-Тян-Шанського.
\end{abstract}

\makeaddressline


\begin{thebibliography}{99}

\bibitem{ReySem}\emph{Рейман А.Г., Семенов-Тян-Шанский М.А.} Лаксово представление со спектральным параметром для волчка Ковалевской и его обобщений // Функц. анализ и его приложения.~-- 1988. -- {\bf 22}, 2. -- С.~87--88.

\bibitem{KhND07}\emph{Харламов М.П.} Критические подсистемы гиростата Ковалевской в двух постоянных полях~// Нелинейная динамика. --  2007. -- {\bf 3}, 3. -- С.~331--348.

\bibitem{PVLect} \emph{Харламов П.В.} Лекции по динамике твердого тела. -- Изд-во НГУ. --  1965. -- 221~с.

\bibitem{Kh34} \emph{Харламов М.П.} Критическое множество и бифуркационная диаграмма задачи о движении волчка Ковалевской в двойном поле // Механика твердого тела.~-- 2004.~-- №~34.~-- С.~47--58.

\bibitem{KhShRCD06} \emph{Kharlamov M.P., Shvedov E.G.} On the existence of motions in the generalized 4th Appelrot class~// Regular and Chaotic Dynamics. --  2006. -- {\bf 11}, 3. -- P.~337-342.

\bibitem{Kh38}\emph{Харламов М.П.} Обобщение 4-го класса Аппельрота: аналитические решения~// Механика твердого тела. -- 2008. -- №~38. -- С.~20--30.

\bibitem{Kh40} \emph{Харламов М.П.} Обобщение 4-го класса Аппельрота: фазовая топология~// Там же. -- 2010. -- №~40. -- С.~21-33.

\bibitem{Kh37}\emph{Харламов М.П.} Особые периодические движения гиростата Ковалевской в двойном поле~// Там же. -- 2007. -- №~37. -- С.~85--96.

\bibitem{KhIISmir}\emph{Харламова И.И., Смирнов Г.Е.} Условия существования периодических движений гиростата Ковалевской в двойном поле~// Там же. -- 2010. -- №~40. -- С.~50--62.

\bibitem{c:main} \emph{Харламов М.П., Рябов П.Е., Савушкин А.Ю., Смирнов Г.Е.} Типы критических точек гиростата Ковалевской в двойном поле~// Там же. -- 2011. -- №~41. -- С.~26-37.

\bibitem{KhRyabUdgu2011} \emph{Харламов М.П., Рябов П.Е.} Диаграммы Смейла-Фоменко и грубые инварианты случая Ковалевской-Яхья~// Вестник УдГУ. -- 2011. -- №~4. -- С.~40-59.

\end{thebibliography}
\end{document}